\documentclass[a4paper,11pt]{article}
\usepackage{jhepmod} % for details on the use of the package, please see the JINST-author-manual
\usepackage{lineno}
\usepackage{bm}
\allowdisplaybreaks[4]

\arxivnumber{1234.56789} % if you have one

\title{Mode analysis of Nambu-Goldstone modes in U(1) charged first-order relativistic viscous hydrodynamics}

\author{Atsuhisa Ota}
\affiliation{Department of Physics and Chongqing Key Laboratory for Strongly Coupled Physics, \\
Chongqing University, Chongqing 401331, People's Republic of China}

\emailAdd{aota@cqu.edu.cn}

\abstract{We conduct a mode analysis of a general $U(1)$-charged first-order relativistic hydrodynamics within the framework of effective field theory for dissipative fluids in flat Minkowski spacetime.
We derive the most general quadratic action for hydrodynamic modes, including stochastic noise, and analyze the corresponding dispersion relations in a consistent gradient expansion.
We argue that spontaneous breaking of spacetime symmetry arises in the presence of a local thermal state specified by a local timelike four-vector.
We demonstrate that hydrodynamical perturbations can be identified as Nambu–Goldstone (NG) modes, analogous to their embedding in global $U(1)$-invariant theories.
We find that frame-invariant combinations of hydrodynamic transport coefficients determine the first-order dispersion relations in the low-energy limit, making the mode analysis manifestly independent of the choice of hydrodynamic frame.
Assuming local Kubo–Martin–Schwinger (KMS) symmetry and unitarity of the underlying UV theory, we show that first-order hydrodynamics is stable if the enthalpy density is positive.

}

\begin{document}
\maketitle
\flushbottom

\section{Introduction}
\label{sec:intro}

Relativistic viscous hydrodynamics is a widely used framework for describing many-body systems whose microscopic interactions are too complex to solve exactly (see Ref.~\cite{Kovtun:2012rj} for a comprehensive review). In this approach, detailed microscopic information is encapsulated in a small number of transport coefficients organized via a derivative expansion. Applications of relativistic hydrodynamics range from cosmology and astrophysics to the physics of ultra-relativistic heavy-ion collisions~\cite{Romatschke:2007mq, Baier:2007ix, Florkowski:2017olj, Jeon:2015dfa, Shibata:2017jyf, Kovtun:2004de, Arnold:2006fz}.

While the formulation of relativistic hydrodynamics is grounded in general principles—such as diffeomorphism invariance and internal symmetries like charge conservation—the inclusion of dissipative effects raises important questions regarding stability and causality~\cite{Hiscock:1983zz, Lindblom:1996gy}. In particular, stability refers to the requirement that perturbations around thermal equilibrium should decay over time, while causality demands that signal propagation be bounded by the speed of light. Both are fundamental requirements for a physically consistent effective theory.

It has been commonly argued that first-order hydrodynamics may violate these conditions. For instance, Ref.~\cite{Hiscock:1985zz} shows that while the Landau frame admits stable hydrodynamic modes in a comoving inertial frame, small deviations from this specific frame can lead to exponential instabilities. Such observations have motivated the development of extended frameworks such as the Müller–Israel–Stewart (MIS) theory~\cite{Israel:1976tn, Israel:1979wp}, which aim to restore stability and causality by incorporating second-order corrections. These frameworks introduce additional transport coefficients and dynamical variables, but also come with increased complexity~\cite{Romatschke:2009im}. However, since different hydrodynamical frame choices in the first order theories correspond to field redefinitions in the thermodynamic variables, physical observables should remain invariant under such transformations. When instabilities arise in specific frames, this may indicate a misapplication of the analysis rather than a fundamental inconsistency in the theory itself.

\medskip

To investigate this point, we study perturbations around hydrodynamic backgrounds within a relativistic EFT framework~\cite{Crossley:2015evo}. A key element of this approach is the identification of dynamical fluctuations as Nambu–Goldstone (NG) modes. Realizing a hydrodynamic background can be interpreted, in a generalized sense, as spontaneous symmetry breaking (SSB): the presence of a fluid rest frame selects a preferred coordinate system. In analogy with phonons in solids, which are NG modes of broken spatial translation symmetry, collective excitations in fluids may also be interpreted as NG modes~\cite{Dubovsky:2011sj, Grozdanov:2013dba}~(See also Ref.~\cite{Huang:2024rml} for a recent proposal about the spontaneous symmetry breaking in hydrodynamics.). This symmetry-based approach has been successfully applied to construct effective actions with manifest covariance, providing a systematic framework to incorporate the noise sector and to impose fluctuation-dissipation relations through the KMS condition at the action level~\cite{Glorioso:2016gsa, Haehl:2014zda}. These features make the mode analysis particularly transparent and robust.

\medskip

We show that some commonly reported instabilities can be traced to mismatches in order counting between the constitutive relations and the perturbative treatment of dynamical modes when solving the constraint equations for the perturbation variables. Similar viewpoints have been expressed in the literature; for example, Ref.~\cite{Kovtun:2012rj} notes that Green’s functions should be used only within the regime of validity of first-order hydrodynamics (see the end of their section~2). Recent works have also re-examined the consistency and well-posedness of first-order theories in a different perspective~\cite{Bemfica:2019cop, Geroch:1990bw,Kovtun:2019hdm}. Here, we aim to clarify this point by performing an explicit analysis with consistent order counting of derivatives in a manifestly covariant framework without hydrodynamical frame dependence, thereby providing a more unified perspective on the stability of relativistic hydrodynamics.

\medskip

This paper is organized as follows. In Sec.~\ref{sec:fast}, we construct the generating functional for a global $U(1)$ theory underlying hydrodynamics by formally integrating out fast modes. The structure of this functional is constrained by unitarity, KMS symmetry, and other properties of the UV theory. 
In Sec.~\ref{fielddef}, we also review the role of field redefinitions and rest-frame ambiguities.

In Sec.~\ref{ngbaction}, we examine the symmetry breaking structure in the presence of a local thermal ensemble. We introduce hydrodynamical modes as the NG modes associated with the spacetime and $U(1)$ symmetries. We derive their quadratic action and eliminate redundant degrees of freedom using the conservation laws.

Section~\ref{sec:md} is devoted to mode analysis in both neutral and $U(1)$-charged fluids. We carefully derive the dispersion relations while keeping track of the gradient expansion. In Sec.~\ref{gfsec}, we compute Green’s functions for the general $U(1)$-charged fluid and examine the analytic structure of the retarded propagators, deriving sound speed bounds and causality constraints.

We conclude with a summary and outlook.  
For the reader’s convenience, we provide a translation table comparing our hydrodynamical notation with that of Ref.~\cite{Kovtun:2019hdm} in App.~\ref{notationdiff}.  
Appendix~\ref{app:usfleqs} collects useful thermodynamic identities, and App.~\ref{appBM} provides a pedagogical review of Brownian motion in a harmonic potential.

\section{Realization of hydrodynamical solutions}
\label{sec:fast}

We adopt the EFT formulation of dissipative hydrodynamics developed in Ref.~\cite{Crossley:2015evo}, which systematically incorporates microscopic constraints such as unitarity and the Kubo-Martin-Schwinger (KMS) condition. While this framework reproduces conventional perturbative results at the linear level (see, e.g., Ref.~\cite{Kovtun:2019hdm}), it offers several additional advantages:

\begin{itemize}
    \item The KMS condition and unitarity are imposed directly on the generating functional, providing a more fundamental alternative to the local second law of thermodynamics. These constraints apply at the level of the effective action itself. When considering a general hydrodynamical frame, the action formalism makes the relationship between the coefficients explicit. 
    
    \item Constraint equations for perturbation variables can be imposed consistently up to the order of noise contributions when derived from an action, as opposed to classical equations of motion.
    
    \item General EFT principles apply: the perturbation variables are understood as nonlinear realizations of spontaneously broken symmetries. This renders the treatment of boost transformations trivial and ensures manifest covariance throughout the analysis.
    
    \item The extension to nonlinear interactions in future work is systematic and straightforward within this formalism.
\end{itemize}

In this section, we demonstrate how a hydrodynamical solution arises from a microscopic theory, providing the groundwork for constructing an EFT of fluctuations around such backgrounds. 

This section serves two purposes: to clarify the internal consistency of our approach by comparing it with existing hydrodynamical frameworks, and to establish the notational and conceptual groundwork for the effective action developed later in this paper. Readers primarily interested in the construction and analysis of the hydrodynamical modes may safely skip to Section~\ref{ngbaction}.

\subsection{Generating functional of UV model}

Let us first consider a UV theory of a complex dynamical variable with a global U(1) symmetry.
A hydrodynamical system can be obtained by coarse-graining over this microscopic system.
Hereafter, we refer to the microscopic degrees of freedom as ``fast modes''.
We couple the fast modes to the spacetime metric $g$ and an external vector field $A$.
The variation of the UV action $S$ with respect to these external fields defines the \textit{microscopic} energy-momentum tensor and conserved current as
\begin{align}
	\hat T^{\mu\nu} \equiv \frac{2}{\sqrt{-\det g}}\frac{\delta S}{\delta g_{\mu\nu}},~\hat J^\mu \equiv \frac{1}{\sqrt{-\det g}} \frac{\delta S}{\delta A_\mu}.
\end{align} 
The corresponding macroscopic observables are then given by
\begin{align}
	T^{\mu\nu} &= \langle \hat T^{\mu\nu}  \rangle,
	\\
	J^{\mu} &= \langle \hat J^{\mu}  \rangle,\label{def:av}
	\\
	W^{\mu\nu\rho\sigma}(x,y)& =\frac{1}{2} \langle \hat T^{\mu\nu}(x)\hat T^{\rho\sigma}(y)\rangle ,
	\\
	X^{\mu\nu\rho}(x,y)& = \frac{1}{2}\langle \hat T^{\mu\nu} (x)\hat J^{\rho} (y)\rangle ,
		\\
	Y^{\mu\rho}(x,y)& =\frac{1}{2} \langle \hat J^{\mu}(x)\hat J^{\rho}(y)\rangle .
\end{align}
These expectation values are defined through the path integral over fast modes using the Schwinger-Keldysh formalism with an appropriate initial state.
At this stage, $T$, $J$, etc., on the left-hand side represent the macroscopic observables obtained by averaging over the microscopic energy-momentum tensor and current under a given initial condition.

We assume that the two-point and higher-order correlation functions are approximately local in space-time:
\begin{align}
	W^{\mu\nu\rho\sigma}(x,y) =W^{\mu\nu\rho\sigma}(x)\frac{\delta(x-y)}{\sqrt{-\det g}}, \label{localnoise}
\end{align}
and similarly for $X$ and $Y$.
Nonlocal correlations will be encoded in long-lived excitations, which will be discussed in the next section.

Given this set of correlation functions, one can reconstruct the generating functional $e^{iI}$ as
\begin{align}
	iI[A,g,\bar A,\bar g] &= \int d^4 x\sqrt{-\det g} \Bigg[ \frac{1}{2}T^{\mu\nu}\bar g_{\mu\nu} + J^{\mu}\bar A_{\mu} 
	\notag
	\\
	&+  \frac{i}{4}\bar g_{\mu\nu}W^{\mu\nu\rho\sigma}\bar g_{\rho\sigma}
	+ i\bar g_{\mu\nu}X^{\mu\nu\rho}\bar A_{\rho}
	+ i\bar A_{\mu}Y^{\mu \rho}\bar A_{\rho}
	+\cdots\Bigg].\label{locaction}
\end{align}
Here, the barred external fields $\bar{g}_{\mu\nu}$ and $\bar{A}_\mu$ represent auxiliary fields in the Keldysh basis of the Schwinger-Keldysh formalism.\footnote{We avoid using the conventional subscript $A$ to prevent confusion with the vector field $A$ and spacetime indices.}
The unbarred external fields couple to physical observables such as $T^{\mu\nu}$ and $J^\mu$.
Due to the assumption~\eqref{localnoise}, the generating functional is local in spacetime.
The correlation functions $T$, $J$, etc.\ are determined by the path integral and characterize the hydrodynamical state of the system.

\medskip
The actual integration is complex in general.
However, one can write the correlation functions based on symmetry structure with undetermined coefficients, which are called constitutive relations.
Let us write the correlation functions for a general U(1)-charged fluid.
A hydrodynamic solution is realized in the presence of a timelike 4-vector $\beta^\mu$, and external fields $g_{\mu\nu}$ and $A_\mu$.
As $\beta^\mu$ splits the external fields into 3+1 form, we can find a convenient set of variables as follows.
The temporal component of the metric is identified with the local temperature $\beta \equiv \sqrt{-g_{\mu\nu}\beta^\mu \beta^\nu}$. We normalize $\beta^\mu$ by $\beta$ and define the fluid 4-velocity $u^\mu \equiv \beta^\mu/\beta$.
The temporal component of the vector field divided by $\beta$ is identified with the chemical potential $\mu\equiv u^\mu A_\mu$.
Finally, the external fields perpendicular to $u^\mu$ are $\gamma_{\mu\nu}\equiv g_{\mu\nu} + u_\mu u_\nu$, and $F_{\mu\nu}=\nabla_\mu A_\nu -\nabla_\nu A_\mu$.
We write $\Psi=\{T,J,W,X,Y\}$ in terms of $\Phi=\{\beta^\mu,g_{\mu\nu},A_\mu\}=\{\beta,u^\mu,\mu,F_{\mu\nu},\gamma_{\mu\nu}\}$.
We consider the most general forms, order by order, in the covariant derivative:
\begin{align}
	\Psi = \sum_{n=0}\Psi_{(n)},
\end{align}
where the subscript $n$ counts the number of covariant derivatives.
Note that $\nabla$ is always projected to $u^\mu$ or $\gamma^{\mu\nu}$.
$F$ is of first order in $\nabla$, and the rest of the variables in $\Phi$ are of zeroth order.

\paragraph{Zeroth order in $\nabla$---.}
In the lowest order, the spacetime indices are only represented by $u^\mu$ and $\gamma^{\mu\nu}$.
The most general $\Psi_{(0)}$ are  
\begin{align}
	T^{\mu\nu}_{(0)} &= \rho u^\mu u^\nu + P\gamma^{\mu\nu},\label{perfect}
	\\
	J^{\mu}_{(0)} &= n u^\mu,\label{perfectJ}
\\
\begin{split}
		W_{(0)}^{\mu\nu\rho\sigma} 
	&=w_1 u^\mu u^\nu u^\rho u^\sigma 
	+ w_2 ( \gamma^{\mu \nu}u^\rho u^\sigma + \gamma^{\rho\sigma}u^\mu u^\nu )
	\\
	&+ w_3 ( \gamma^{\mu \rho}u^\nu u^\sigma + \gamma^{\nu \rho}u^\mu u^\sigma +\gamma^{\mu \sigma}u^\nu u^\rho+\gamma^{\nu \sigma}u^\mu u^\rho)
	\\ 
	&+ w_4 \gamma^{\mu\nu}\gamma^{\rho\sigma}
	+ w_5\left( \gamma^{\mu\rho}\gamma^{\nu\sigma} 
	+\gamma^{\nu\rho}\gamma^{\mu\sigma} 
	-\frac{2}{3}\gamma^{\mu\nu}\gamma^{\rho\sigma}\right ).
\end{split}
\label{def:W0}
\\
\begin{split}
		X_{(0)}^{\mu\nu\rho} 
	&=\beta x_1 u^\mu u^\nu u^\rho  
	+ \beta x_2  \gamma^{\mu \nu}u^\rho  
	+ \beta x_3 ( \gamma^{\mu \rho}u^\nu  + \gamma^{\nu \rho}u^\mu  ).
\end{split}
\label{def:Wd}
\\
\begin{split}
		Y_{(0)}^{\mu \rho } 
	&=\beta^2 y_1 u^\mu  u^\rho  
	+ \beta^2 y_2  \gamma^{\mu \rho}.
\end{split}
\label{def:W0}
\end{align}
Note that we assumed Eq.~\eqref{localnoise}.
The index symmetry in correlation functions implies $W^{\mu\nu\rho\sigma}=W^{\nu\mu\rho\sigma}=W^{\rho\sigma\mu\nu}$, $X^{\mu\nu\rho}=X^{\nu\mu\rho}$, and $Y^{\mu\nu}=Y^{\nu\mu}$.
The expansion coefficients $\rho$, $P$, $n$, $w$, $x$ and $y$ are functions of $\beta$ and $\mu$, which are undetermined within the effective theory.
The mass dimensions of the coefficients are given as $[T]=M^4$, $[J]=M^3$, $[W]=M^4$, $[X]=M^3$, and $[Y]=M^2$.

\paragraph{First order in $\nabla$---.}

The relevant first-order derivative corrections are  
\begin{align}
\begin{split}
	T^{\mu \nu}_{(1)} &=  \rho_{(1)} u^\mu u^\nu +P_{(1)}\gamma^{\mu\nu} + q_{(1)\sigma} ( u^\mu\gamma^{\nu\sigma} + u^\nu\gamma^{\mu\sigma})
	- \Sigma_{(1)}^{\mu\nu}
\end{split}
	\label{emtt1grad},
	\\
	\begin{split}
	J^{\mu }_{(1)} &= n_{(1)}u^\mu + \gamma^{\mu\nu}j_{(1)\nu} ,
\end{split}
\end{align}
where we define the following first-order coefficients:
\begin{align}
	\rho_{(1)} &= \beta^{-1}\epsilon u^\rho \nabla_\rho \beta  - \lambda_1 \nabla_\rho u^\rho - \nu_4 u^\rho \nabla_\rho (\beta\mu) \label{rho1} ,
	\\
	P_{(1)} &= \beta^{-1} {\lambda_2} u^\rho \nabla_\rho \beta  - \zeta \nabla_\rho u^\rho  -  \nu_5 u^\rho \nabla_\rho (\beta \mu) ,
	\\
	n_{(1)} & =  \nu_1 u^\rho \nabla_\rho  \beta  - \beta \nu_2 \nabla_\rho u^\rho -  \beta \nu_3 u^\rho \nabla_\rho(\beta   \mu),
\\
	q_{(1)\mu} &= \beta^{-1}\kappa_1 \nabla_\mu  \beta - \kappa_2  u^\rho\nabla_\rho  u_\mu - \tau_3 \nabla_\mu (\beta \mu) - \beta \tau_4 u^\rho  F_{\rho\mu},\label{q1}
	\\
	j_{(1)\mu} &= \tau_1 \nabla_\mu  \beta -  \beta \tau_2 u^\rho\nabla_\rho  u_\mu  -  \beta \chi_1  \nabla_\rho  (\beta \mu) - \beta^2 \chi_2 u^\rho F_{\rho\mu},
\\
	\Sigma_{(1)}^{\mu\nu} &= \eta \gamma^{\mu\alpha}\gamma^{\nu\beta}\left(\nabla_\beta u_\alpha  + \nabla_\alpha u_\beta - \frac{2}{3}\gamma_{\alpha\beta}\nabla_\gamma u^\gamma \right).\label{deffirstsigma}
\end{align}
These are the most general forms allowed by general covariance.
All first-order coefficients have mass dimension three.
The effective theory is organized as an expansion in $\varepsilon \equiv (\rho+P)^{-1}v \nabla$, with viscosity coefficients $v = \epsilon, \lambda, \zeta$, etc.
Readers can find a table summarizing the notational differences from a seminal work~\cite{Kovtun:2019hdm} in Appendix~\ref{notationdiff}.

\subsection{Constraints on the generating functional}

The above considerations align with the traditional derivative expansion of first-order hydrodynamics. 
With this setup, one can write the hydrodynamical equations and discuss the dynamics of $T^{\mu\nu}$ and $J^\mu$. 
Moreover, by constructing the entropy current, the local second law of thermodynamics introduces additional constraints on the hydrodynamical coefficients.

We reconstruct the effective action from these constitutive relations. 
The alternative to the local second law is the imposition of the unitarity and KMS conditions on the generating functional. 
With these conditions, the effective action is fixed up to the noise order, i.e., $W$, $X$, and $Y$ are related to $T$ and $J$. 

Eq.~\eqref{locaction} is obtained by integrating out the fast modes in the UV theory using the Schwinger-Keldysh formalism. 
In the operator formalism, the path integral can be written as
\begin{align}
	e^{iI} = {\rm Tr}[\hat D \hat U^\dagger_2 \hat U_1],\label{defgf}
\end{align}
where $\hat D$ is the density operator, and $\hat U_i$ is the unitary operator defined on the $i$-th path of the closed time path~(CTP).
External fields are suppressed for notational simplicity.
The physical and noise fields in the Keldysh basis are defined as the average and difference of those defined on the first and second contours of the CTP, namely $A\equiv (A_1+A_2)/2$ and $\bar A\equiv A_1-A_2$.
Consistency with the UV theory requires the generating functional to satisfy the following conditions~\cite{Crossley:2015evo, Glorioso:2017fpd}:

\paragraph{Conjugate condition.} The Hermiticity of the density operator $\hat D=\hat D^\dagger$ implies
\begin{align}
	{\rm Tr}[\hat D \hat U^\dagger_2 \hat U_1]^\dagger = {\rm Tr}[\hat U^\dagger_1 \hat U_2 \hat D^\dagger]= {\rm Tr}[\hat D \hat U^\dagger_1 \hat U_2 ].\label{conjugagecond}
\end{align}	 
In the Keldysh basis, this is equivalent to
\begin{align}
	-I^*[g,A,\bar g,\bar A] = I[g,A,-\bar g,-\bar A].
\end{align}

\paragraph{Unitarity condition.} The Schwinger-Keldysh path integral yields a trivial result when $\hat U_1 = \hat U_2$. In the Keldysh basis, this corresponds to the absence of noise:
\begin{align}
	I[g,A,\bar g=0,\bar A=0]= 0.\label{firstunitary}
\end{align}	
The trace inequality $|{\rm Tr}[\hat D \hat U_2^\dagger \hat U_1]|\leq 1$ further implies~(see also Appendix of Ref.~\cite{Ota:2024mps})
\begin{align}
	{\rm Im}~I\geq 0.\label{secondunitary}
\end{align}

\paragraph{Local KMS condition.} 
The local KMS condition arises as a remnant of microscopic time-reversal symmetry in a thermal ensemble and plays a crucial role in constraining the structure of the effective action.

Let us first consider the condition for microscopic time-reversal symmetry. Define the time-reversed evolution operators as $\hat{\mathcal U}_i = \mathcal T \hat U_i \mathcal T^\dagger$, where $\mathcal T$ is a time-reversal operator. If the initial density matrix is invariant under time reversal, i.e., $\hat D = \mathcal T^\dagger \hat D \mathcal T$, then the following identity holds:
\begin{align}
	{\rm Tr}\left[\hat D \hat U_1 \hat U_2^{-1}\right] = {\rm Tr}\left[\hat D \hat{\mathcal U}_2^{-1} \hat{\mathcal U}_1\right].\label{deftimerevs}
\end{align}
The left-hand side represents a CTP evolution ending in state $\hat D$, while the right-hand side corresponds to a time-reversed evolution of all operators. This microscopic symmetry implies a constraint on the generating functional, which underlies the so-called local KMS condition.

To understand its concrete implications, consider a system in thermal equilibrium with inverse temperature $\beta$ and chemical potential $\mu$. The corresponding density operator is
\begin{align}
	\hat D = \frac{1}{Z} e^{-\beta (\hat H - \mu \hat Q)}, \quad Z = {\rm Tr}[e^{-\beta (\hat H - \mu \hat Q)}].
\end{align}
For simplicity, we focus on the Hamiltonian part and assume $[\hat H, \hat Q] = 0$. Since $e^{-\beta \hat H}$ generates imaginary-time translations, the trace
\begin{align}
	{\rm Tr}[e^{-\beta \hat H} \hat U_2^{-1} \hat U_1]
\end{align}
can be rewritten by inserting imaginary-time translation operators before and after $\hat U_1$ and $\hat U_2^{-1}$:
\begin{align}
	= {\rm Tr}[e^{-\beta \hat H} \hat U_2^{-1} e^{(\beta - d)\hat H} e^{-\beta \hat H} e^{d \hat H} \hat U_1 e^{-d \hat H} e^{d \hat H}],
\end{align}
where $d \in [0,\beta)$ is arbitrary. By cyclicity of the trace, this becomes
\begin{align}
	{\rm Tr}[e^{-\beta \hat H} \hat U_{1\uparrow} \hat U_{2\downarrow}^{-1}],\label{ndnoisetime}
\end{align}
with 
\begin{align}
	\hat U_{1\uparrow} &= e^{d \hat H} \hat U_1 e^{-d \hat H},\label{imup}\\
	\hat U_{2\downarrow}^{-1} &= e^{-(\beta-d)\hat H} \hat U_2^{-1} e^{(\beta - d) \hat H}.\label{imdown}
\end{align}

Now, suppose the effective action is written in terms of external sources $A_i$ on the two legs of the CTP~($i=1,2$). The above imaginary-time translations correspond to a shift of these sources. For $d = \beta/2$, the infinitesimal form of this transformation becomes:
\begin{align}
	e^{\pm \frac{i}{2}\hbar \pounds_\beta} A_i = A_i \pm \frac{i\hbar}{2} \pounds_\beta A_i + \mathcal O(\hbar^2),
\end{align}
where $\pounds_\beta$ is the Lie derivative along the thermal vector $\beta^\mu$. In the Keldysh basis,
\begin{align}
	A &\to A + \mathcal O(\hbar^2),\\
	\bar A &\to \bar A + i\hbar \pounds_\beta A + \mathcal O(\hbar^3),
\end{align}
i.e., the physical field $A$ remains unchanged at leading order, while the noise field $\bar A$ acquires a shift proportional to $\pounds_\beta A$.

This shift reflects a symmetry of the effective action under a combination of imaginary-time translations and time reversal, which is called the local KMS symmetry. It provides a $Z_2$-type constraint on the effective action that becomes manifest in the semiclassical ($\hbar \to 0$) limit. This symmetry is instrumental in deriving fluctuation-dissipation relations and ensuring the thermal consistency of dissipative dynamics.

While we have focused here on time-reversal symmetry $\mathcal T$, the same argument can be extended to $\mathcal P\mathcal T$ or $\mathcal C\mathcal P\mathcal T$ symmetry, depending on the discrete symmetries of the underlying UV theory.

\medskip
Eq.~\eqref{locaction} satisfies the conjugate condition~\eqref{conjugagecond} and the first unitarity condition~\eqref{firstunitary} by construction.
In the next section, we impose the second unitarity condition~\eqref{secondunitary} and the local KMS condition to further constrain the undetermined coefficients in the constitutive relations.

\subsection{Local KMS condition}

In this subsection, we derive a consequence of the KMS condition.
Consider a local KMS transformation, which consists of a $\mathcal{P} \mathcal{T}$ transformation followed by an infinitesimal noise time diffeomorphism generated by $i\hbar \beta^\mu$.
A covariant derivative and the coordinate system are $\mathcal{P} \mathcal{T}$-odd, whereas the 4-velocity $u^\mu$ and the external fields $A_\mu$ and $g_{\mu\nu}$ are $\mathcal{P} \mathcal{T}$-even~(see Appendix C of Ref.~\cite{Liu:2018kfw}).
Note that $u^\mu \partial_\mu$ is $\mathcal{P} \mathcal{T}$-odd, implying that $u^\mu$ itself is even.

Since $u^\mu$ is $\mathcal{P} \mathcal{T}$-even and $\nabla$ is $\mathcal{P} \mathcal{T}$-odd, the zeroth-order quantities $T_{(0)}^{\mu\nu}$ and $J_{(0)}^\mu$ are $\mathcal{P} \mathcal{T}$-even, while the first-order corrections $T_{(1)}^{\mu\nu}$ and $J_{(1)}^\mu$ are $\mathcal{P} \mathcal{T}$-odd.
Thus, the $\mathcal{P} \mathcal{T}$ transformations for these composite quantities are not trivially dictated by coordinate transformation, much like the parity transformation for the electric and magnetic fields.

The transformation of Eq.~\eqref{locaction} under the local KMS symmetry is:
\begin{align}
	T^{\mu\nu}\bar g_{\mu\nu} &\to T_{(0)}^{\mu\nu}\bar g_{\mu\nu} - T_{(1)}^{\mu\nu}\bar g_{\mu\nu} + i\hbar T_{(0)}^{\mu\nu} \pounds_\beta g_{\mu\nu}  - i\hbar T_{(1)}^{\mu\nu} \pounds_\beta g_{\mu\nu} ,
		\\
	J^\mu  \bar A_\mu &\to J^\mu_{(0)}  \bar A_\mu -  J^\mu_{(1)}  \bar A_\mu + J^\mu_{(0)}i\hbar   \pounds_\beta  A_\mu  - J^\mu_{(1)}  i\hbar   \pounds_\beta  A_\mu,
	\\
	iW_{(0)}^{\mu\nu\alpha \beta}\bar g_{\mu\nu}\bar g_{\alpha\beta} &\to iW_{(0)}^{\mu\nu\alpha \beta}\left(\bar g_{\mu \nu}+ i\hbar \pounds_\beta g_{\mu\nu}\right)\left(\bar g_{\alpha \beta}+ i\hbar \pounds_\beta g_{\alpha\beta}\right) ,
	\\
	i X^{\mu \nu \rho }_{(0)}\bar g_{\mu \nu}\bar A_\rho &\to i X^{\mu \nu \rho }_{(0)}\left(\bar g_{\mu \nu}+ i\hbar \pounds_\beta g_{\mu\nu}\right)\left(\bar A_\rho + i\hbar \pounds_\beta A_\rho\right),
		 \\
	i Y^{\mu  \rho }_{(0)}\bar A_{\mu }\bar A_\rho &\to i Y^{\mu  \rho }_{(0)}(\bar A_{\mu }+i\hbar \pounds_\beta A_\mu)(\bar A_\rho+i\hbar \pounds_\beta A_\rho).
\end{align}
We treat $\beta^\mu$ as finite but expand in small $\hbar$, keeping only the leading-order terms.
Note that noise fields are counted as $\mathcal{O}(\hbar)$ since they vanish on the classical path.

The variation of the generating functional under this transformation is
\begin{align}
	\delta_{\rm KMS} I=&  \int d^4 x \sqrt{-\det g}\Bigg[\frac{1}{2}i\hbar T_{(0)}^{\mu\nu} \pounds_\beta g_{\mu\nu} + i\hbar J^\mu_{(0)}   \pounds_\beta  A_\mu
	\notag 
	\\
	&
	- \left(  T_{(1)}^{\mu\nu}	+ \frac{\hbar}{2} W_{(0)}^{\mu\nu\alpha \beta}\pounds_\beta g_{\alpha\beta}
	+\hbar X^{\mu \nu \rho }_{(0)} \pounds_\beta A_\rho \right)\left( \bar g_{\mu\nu} + \frac{i}{2}\hbar \pounds_\beta g_{\mu\nu}\right)
	\\
	&
	-2 \left( J^\mu_{(1)}   + \hbar  Y^{\mu  \rho }_{(0)} \pounds_\beta A_\rho
	+\frac{\hbar}{2} X^{\rho \nu \mu }_{(0)}  \pounds_\beta g_{\rho\nu}\right)\left(\bar A_\mu + \frac{i}{2}\hbar      \pounds_\beta  A_\mu \right)\bigg].
\end{align}

The first line can be simplified to a total derivative.
We first compute
\begin{align}
	\frac{1}{2} T_{(0)}^{\mu\nu} \pounds_\beta g_{\mu\nu} &= \left( (\rho+P)u^\mu u^\nu  + Pg^{\mu\nu} \right) \nabla_\mu \beta_\nu
	\notag 
	\\
	&= -\left[ \beta \left(\frac{\partial P}{\partial \beta}\right)_{\mu} +\rho+P\right]u^\mu   \nabla_\mu \beta  -   \beta \left(\frac{\partial P}{\partial \mu}\right)_{\beta} u^\mu \nabla_\mu \mu + \nabla_\mu (P \beta^\mu)
\end{align}
and
\begin{align}
	J^\mu_{(0)}   \pounds_\beta  A_\mu 
	= n  \mu  u^\mu  \nabla_\mu  \beta  + n \beta^\rho  \nabla_\rho \mu .
\end{align}

Combining these results with the thermodynamical relations in App.~\ref{app:usfleqs}, we obtain
\begin{align}
	\frac{1}{2}T_{(0)}^{\mu\nu} \pounds_\beta g_{\mu\nu} +J^\mu_{(0)}   \pounds_\beta  A_\mu 
	=  \nabla_\mu (P \beta^\mu).
\end{align}

If the covariant fluctuation-dissipation relations (FDRs)
\begin{align}
	T_{(1)}^{\mu\nu} +\hbar  X^{\mu \nu \rho }_{(0)} \pounds_\beta A_\rho + \frac{\hbar}{2} W_{(0)}^{\mu\nu\alpha \beta}\pounds_\beta g_{\alpha\beta}
	 &= 0,\label{fdrT}
	\\
	J^\mu_{(1)}   +\hbar  Y^{\mu  \rho }_{(0)} \pounds_\beta A_\rho
	+\frac{\hbar}{2} X^{\rho \nu \mu }_{(0)}  \pounds_\beta g_{\rho\nu}&=0,\label{fdrJ}
\end{align}
are satisfied, the variation becomes
\begin{align}
	\delta_{\rm KMS} I =& \int d^4 x \sqrt{-\det g} \nabla_\mu (P\beta^\mu) + \mathcal O(\hbar^3).
\end{align}

The boundary term does not affect the variation of the generating functional with respect to external sources and can be absorbed into its normalization.
Under the local KMS condition, noise terms consistently scale with $\hbar$, reflecting the emergence of classical statistical noise from quantum fluctuations.
At this stage, we note that the local KMS condition can be generalized to a local vector $\beta^\mu(x)$. If the generating functional satisfies noise diffeomorphism invariance and micro-time reversality, one may introduce a local vector $\beta^\mu(x)$ that generates the thermodynamic relations and FDRs.
This constitutes an implicit introduction of a local Gibbs distribution.
The $Z_2$ transformation is expressed in a covariant manner.

\subsection{Fluctuation-dissipation relations}

The FDRs~\eqref{fdrT} and \eqref{fdrJ} constrain the effective action.
The computation of the Lie derivatives of the metric and vector field with respect to $\beta^\mu$ is straightforward.
We find
\begin{align}
	\frac{\rho_{(1)}}{\beta \hbar } &= w_1  \beta^{-1} u^\rho  \nabla_\rho \beta
	 -w_2  \nabla_\rho u^\rho -  x_1  u^\rho  \nabla_\rho(\beta \mu), \\
	\frac{P_{(1)}}{\beta \hbar} &= w_2  u^\rho  \beta^{-1} \nabla_\rho \beta - w_4 \nabla_\rho u^\rho - x_2 u^\rho \nabla_\rho(\beta \mu), \\
	\frac{n_{(1)}}{ \beta \hbar } &=  x_1  u^\rho    \nabla_\rho \beta 
	- x_2 \beta  \nabla_\rho u^\rho - y_1 \beta  u^\rho  \nabla_\rho(\beta \mu ), \\
	\frac{q_{(1)\sigma}}{\beta \hbar } &= w_3 \left( \beta^{-1}\nabla_\sigma\beta- u^\rho\nabla_\rho u_\sigma \right) -   x_3 \left( \beta u^\sigma    F_{\sigma \rho} + \nabla_\sigma (\beta \mu)\right), \\
	\frac{j_{(1)\mu}}{\beta \hbar } &= x_3 \left( \nabla_\mu \beta -  \beta u^\rho  \nabla_\rho u_\mu\right)
	-
	 y_2   \left(\beta^2    u^\nu F_{\nu\mu}  +  \beta\nabla_\mu(\beta \mu )\right) ,
\end{align}
Then, the FDRs for the hydrodynamical coefficients are obtained as~\cite{Crossley:2015evo}
\begin{align}
	w_1 &= \frac{\epsilon}{\beta \hbar}, \quad w_2 = \frac{\lambda_1}{\beta \hbar} = \frac{\lambda_2}{\beta \hbar}, \quad w_3 = \frac{\kappa_1}{\beta \hbar} = \frac{\kappa_2}{\beta \hbar}, 
	w_4 = \frac{\zeta}{\beta \hbar}, \quad w_5 = \frac{\eta}{\beta \hbar}, \\
	x_1 &= \frac{\nu_1}{\beta \hbar} = \frac{\nu_4}{\beta \hbar}, \quad x_2 = \frac{\nu_2}{\beta \hbar} = \frac{\nu_5}{\beta \hbar}, \quad x_3 = \frac{\tau_1}{\beta \hbar} = \frac{\tau_2}{\beta \hbar} = \frac{\tau_3}{\beta \hbar} = \frac{\tau_4}{\beta \hbar}, \\
	y_1 &= \frac{\nu_3}{\beta \hbar}, \quad y_2 = \frac{\chi_1}{\beta \hbar} = \frac{\chi_2}{\beta \hbar}.
\end{align}
For thermal states, we use the following conventional parameters:
\begin{align}
\lambda \equiv \lambda_n, \quad \kappa \equiv \kappa_n, \quad \tau \equiv \tau_n, \quad \chi \equiv \chi_n. \label{newparams}
\end{align}
Note that we have evaluated the covariant FDRs, and thus the above FDRs hold in a general curved spacetime.

\subsection{Unitarity constraints}

The unitarity condition further constrains the noise coefficients.
Consider the 3+1 decomposition of the noise external fields:
\begin{align}
	\bar {N} &\equiv  u^\mu u^\nu \bar g_{\mu\nu}, \\
	\bar {N}_{\mu} &\equiv u^\rho  \gamma_{\mu}{}^\nu\bar g_{\rho\nu}, \\
	\bar \gamma_{\mu\nu} &\equiv   \gamma_{\mu}{}^\rho \gamma_{\nu}{}^\sigma \bar g_{\rho\sigma}, \\
	\bar \gamma &\equiv \gamma^{\mu\nu} \bar \gamma_{\mu\nu}, \\
	\bar \gamma_{\perp \mu\nu} &\equiv \bar \gamma_{\mu\nu} - \frac{\bar \gamma}{3} \gamma_{\mu\nu}, \\
	\bar A &\equiv u^\rho  \bar A_{\rho}, \\
	\bar {A}_{\perp \mu} &\equiv \gamma_{\mu\nu}  \bar A^{\nu}. 
\end{align}
Then, the noise part of the effective action is written as
\begin{align}
 {\rm Im}~ I &= \int d^4 x \sqrt{-\det g} \frac{1}{\beta}\Bigg[\frac{\epsilon}{4}\bar N^2 + \frac{\zeta}{4}\bar \gamma^2 + \frac{\lambda}{2}\bar N \bar \gamma + \frac{\eta}{2}\bar \gamma_{\perp \mu \nu}\bar \gamma_{\perp}^{\mu\nu}  + \kappa \bar N_\mu\bar N^\mu  
 \notag \\
 &+ \nu_1 \bar A \bar N + \nu_2 \bar A \bar \gamma  + \nu_3 \bar A^2 + \chi \bar A_{\perp \mu} \bar A_{\perp}^{ \mu} + 2\tau \bar A_{\perp \mu} \bar N^\mu \Bigg]. \label{noiseactiong}	
\end{align}
For arbitrary external fields, the second unitarity condition~\eqref{secondunitary} is satisfied.
Thus, the following parameter space is allowed~\cite{Crossley:2015evo}:
\begin{align}
	\eta &\geq 0, \\
	\kappa x^2 + 2\tau xy + \chi y^2 &\geq 0, \label{positivdefeplam2} \\
	\epsilon x^2 + \zeta y^2 + \nu_3 z^2 + 2 \lambda xy + 2 \nu_1 yz + 2 \nu_2 zx &\geq 0, \label{positivdefeplam}  
\end{align}
where $x$, $y$, and $z$ are arbitrary real numbers.

\paragraph{Section summary.} 
We have constructed the generating functional for a general system with a $U(1)$ symmetry, incorporating all terms up to the noise order that are allowed by general covariance. The only fundamental input in this formulation is a four-vector $\beta^\mu$, along with the external sources $A_\mu$ and $g_{\mu\nu}$. The vector $\beta^\mu$ naturally induces a $3+1$ decomposition of the external fields. While we refer to the projections of these fields along $\beta^\mu$ as the temperature and chemical potential, thermodynamics are not assumed \textit{a priori}—$\beta^\mu$ merely characterizes the flow of the coarse-grained system. Hence, at this stage, arbitrary first order non-equilibrium many-body system characterized by a 4-vector is expressed in this form.

By imposing the local KMS condition—as a remnant of microscopic time-reversal symmetry and grand canonical ensemble—thermodynamic relations emerge, and the first-order transport coefficients become related to the corresponding noise coefficients. The unitarity condition on the noise sector is then reformulated as constraints on the dissipative coefficients, such as viscosities. In traditional hydrodynamics, these constraints are usually derived from the local second law of thermodynamics imposed phenomenologically; in contrast, our approach derives them directly from the structure of the effective action.

At the outset, the system contains three thermodynamic variables, seventeen first-order dissipative coefficients, and ten zeroth-order noise coefficients. The local KMS condition imposes a thermodynamic relation that allows $\rho$ and $n$ to be expressed in terms of $P$. Its extension to first order—the fluctuation-dissipation relation (FDR)—further relates the seventeen dissipative coefficients to the ten noise coefficients. As a result, the total number of independent parameters is reduced from thirty to eleven~\footnote{More rigorously, the equation of state is necessary to fully determine the relationship in thermodynamical variables.}.

Up to this point, the effective action has been constrained solely by physical principles. Nevertheless, it still includes redundancies associated with field redefinitions—purely unphysical degrees of freedom—which must be properly fixed. These will be addressed in the next section.

\section{Frame redundancy and hydrodynamical invariants}
\label{fielddef}

When we go to first order theories, there are inherent ambiguities in the definition of $\beta^\mu$ at first order. These ambiguities are traditionally referred to as hydrodynamical frame transformations, which correspond to redefinitions of the hydrodynamic fields that leave the physical content unchanged. Since these transformations merely represent different parametrizations of the same underlying dynamics, the generating functional must remain invariant under them. Consequently, physically meaningful results, such as mode analyses, should be expressed in terms of hydrodynamical frame invariants.

In this section, we clarify how the first-order transport coefficients transform under field redefinitions and identify the combinations that remain invariant.

\subsection{Field redefinition}

As summarized in the previous section, $\beta^\mu$ introduces a rich structure in the constitutive relations. When expanding these relations to first order in derivatives, one may incorporate the expansion parameters into $\beta^\mu$ to reduce the number of independent coefficients. This redundancy is associated with $\beta^\mu$ itself, rather than with the external fields.

As previously noted, $\mu$ and $\beta$ are projections of the external fields onto $\beta^\mu$. Therefore, we cannot redefine both independently. Once we fix $\beta^\mu$ by choosing $\delta \beta^\mu = u^\mu \delta \beta  + \beta \delta u^\mu$, there is no remaining freedom in $\mu$. Alternatively, we may fix $\delta \beta = 0$ and allow $\delta \mu \neq 0$.

As a general choice, one often considers a redefinition of the hydrodynamical fields~\cite{Kovtun:2012rj}:
\begin{align}
	\beta &\to \beta + \delta \beta, \\
	u^\mu &\to u^\mu + \delta u^\mu, \\
	\mu &\to \mu + \delta \mu.
\end{align}
$\delta \beta$, $\delta u^\mu$, and $\delta \mu$ are generally written as
\begin{align}
	\delta \beta &= \beta^{-1}\epsilon_\beta  u^\rho \nabla_\rho \beta  - \lambda_\beta  \nabla_\rho u^\rho - \nu_\beta u^\rho \nabla_\rho (\beta\mu), \\
	\delta \mu &= \beta^{-1}\epsilon_\mu  u^\rho \nabla_\rho \beta  - \lambda_\mu  \nabla_\rho u^\rho - \nu_\mu u^\rho \nabla_\rho (\beta\mu), \\
	\delta u_{\mu} &= \beta^{-1}\kappa_\beta \nabla_\mu  \beta - \kappa_u  u^\rho\nabla_\rho  u_{\mu} - \tau_{\beta \mu} \nabla_\mu (\beta \mu) - \beta \tau_{F} u^\rho  F_{\rho\mu},
\end{align}
with $\delta u^\mu = \gamma^{\mu\rho}\delta u_\rho$. While ten parameters appear in the expressions above, not all are independent, as discussed earlier.

The redefinition of the hydrodynamical solutions propagates to the hydrodynamical coefficients as follows.
First, the zeroth-order coefficients are expanded as
\begin{align}
	\delta \rho &= \rho_\beta \delta \beta + \rho_\mu \delta \mu, \\
	\delta P &= P_\beta \delta \beta + P_\mu \delta \mu, \\
	\delta n &= n_\beta \delta \beta + n_\mu \delta \mu.  
\end{align}
Since $\delta \beta$, $\delta u^\mu$, and $\delta \mu$ are first-order terms, $\delta \rho$, $\delta P$, and $\delta n$ introduce the transformation for the first-order coefficients:
\begin{align}
	\delta \epsilon &= \rho_\beta \epsilon_\beta + \rho_\mu \epsilon_\mu ,\label{kmsfdef:begin} \\
	\delta \lambda_1 &= \rho_\beta \lambda_\beta + \rho_\mu \lambda_\mu ,\label{kmsconsistent1} \\
	\delta \nu_4 &= \rho_\beta \nu_\beta + \rho_\mu \nu_\mu  ,\label{kmsconsistent2} \\
	\delta \lambda_2 &= P_\beta \epsilon_\beta + P_\mu \epsilon_\mu , \\
	\delta \zeta & = P_\beta \lambda_\beta + P_\mu \lambda_\mu, \\
	\delta \nu_5 & = P_\beta \nu_\beta  + P_\mu \nu_\mu, \\
	\delta \nu_1 & = n_\beta \epsilon_\beta + n_\mu \epsilon_\mu, \\
	\delta \nu_2 &= n_\beta \lambda_{\beta}  + n_\mu \lambda_{\mu},\label{kmsconsistent3} \\
	\delta \nu_3 &= n_\beta \nu_\beta  + n_\mu \nu_\mu.\label{kmsfdef:end}
\end{align}
Similarly, we have
\begin{align}
	\delta q_\mu &= (\rho +P)\delta u^\mu, \\
	\delta j_\mu &= n \delta u^\mu. 
\end{align}
Then, we find
\begin{align}
	\delta \kappa_1 &=(\rho +P)\kappa_\beta, \\
	\delta \kappa_2 &=(\rho +P)\kappa_{u}, \\
	\delta \tau_3 &=(\rho +P)\tau_{\beta \mu}, \\
	\delta \tau_4 &=(\rho +P)\tau_{F}, \\
	\delta \tau_1 &=\frac{n}{\beta}\kappa_\beta, \\
	\delta \tau_2 &=\frac{n}{\beta}\kappa_{u}, \\
	\delta \chi_1 &=\frac{n}{\beta}\tau_{\beta \mu}, \\
	\delta \chi_2 &=\frac{n}{\beta}\tau_{F}.
\end{align}
The field redefinition does not change the leading-order noise coefficients $x_n$, $y_n$, and $w_n$ as $\delta \beta$, $\delta \mu$, and $\delta u^\mu$ are first-order terms.

Any field redefinition in the KMS-symmetric generating functional generally leads to a violation of the \textit{explicit} form of the KMS condition with respect to the redefined $\beta^\mu$. 
This implies that the new $\beta^\mu$ can no longer be interpreted as the genuine thermodynamic temperature vector that enters the grand canonical ensemble at first order.%
\footnote{We emphasize that only the explicit form of the KMS condition is not preserved; the initial thermal state of the generating functional remains unchanged. 
The KMS symmetry is still satisfied in a modified, implicit form, as discussed in Ref.~\cite{Glorioso:2017fpd}.}
This trade-off allows one to reduce the number of hydrodynamical coefficients at the expense of losing manifest fluctuation-dissipation relations.
When maximally eliminating the coefficients with $\rho_{(1)}=n_{(1)}=0$ and $q^\mu=0$ the remaining parts are written as
\begin{align}
	P_{(1)} &= \beta^{-1} \tilde \lambda u^\rho \nabla_\rho \beta  - \tilde \zeta \nabla_\rho u^\rho  -  \tilde \nu u^\rho \nabla_\rho (\beta \mu) ,\label{Plan}
	\\
	j_{(1)\mu} &= \tilde \tau_1  \nabla_\mu  \beta -  \tilde \tau_2 \beta u^\rho\nabla_\rho  u_\mu   - \tilde  \chi_1  \beta  \nabla_\rho  (\beta \mu) -\tilde  \chi_2 \beta^2  u^\rho F_{\rho\mu},\label{Jlan}
\end{align}
where we defined the following hydrodynamical frame invariants:
\begin{align}
	\tilde \lambda  &= \lambda_2 - \left(\frac{\partial P}{\partial \rho}\right)_n \epsilon  - \left(\frac{\partial P}{\partial n}\right)_\rho \nu_1,
	\\
	\tilde \zeta  &= \zeta - \left(\frac{\partial P}{\partial \rho}\right)_n \lambda_1  - \left(\frac{\partial P}{\partial n}\right)_\rho \nu_2,
	\\
	\tilde \nu  &= \nu_5 - \left(\frac{\partial P}{\partial \rho}\right)_n \nu_4  - \left(\frac{\partial P}{\partial n}\right)_\rho \nu_3,\label{nu2inv}
	\\
	\tilde \tau_1 & = \tau_1 -\frac{n}{\beta(\rho+P)} \kappa_1 ,
	\\
	\tilde \tau_2 & = \tau_2 -\frac{n}{\beta(\rho+P)} \kappa_2 ,
	\\
	\tilde  \chi_1 & = \chi_1 - \frac{n}{\beta(\rho+P)}\tau_3,
	\\
	\tilde  \chi_2 & = \chi_2 - \frac{n}{\beta(\rho+P)}\tau_4.
\end{align}
When starting from the most general U(1) charged hydrodynamics with the KMS condition, $\nu_4=\nu_1$, $\nu_5=\nu_2$, and Eq.~\eqref{newparams} are satisfied~(see app.~\ref{KMSsummary}.).
In this case, we reduce the number of vector frame invariants to $\tilde \tau = \tilde \tau_i$ and $\tilde \chi = \tilde \chi_i$.
%The KMS condition is not satisfied for the redefined hydrodynamical coefficients explicitly, so we can no longer constrain them.

\subsection{Hydrodynamical on-shell transformation}

To derive Eqs.~\eqref{Plan} and~\eqref{Jlan}, we have exhausted the freedom of hydrodynamical frame transformations. The transport coefficients with a tilde are hydrodynamical frame invariants, so we cannot further eliminate them by the hydrodynamical frame transformation. However, when evaluated on the lowest-order hydrodynamical solutions, one can further simplify the coefficients~\cite{Kovtun:2012rj}.

The lowest-order hydrodynamical equations are:
\begin{align}
	\nabla_\mu (n u^\mu) &= 0, \\
	u_\nu \nabla_\rho ( \rho u^\rho u^\nu + P \gamma^{\rho\nu}) &= 0, \\
	\gamma_{\mu\nu} \nabla_\rho ( \rho u^\rho u^\nu + P \gamma^{\rho\nu}) +  \gamma_{\mu\nu} n u_\rho F^{\rho \nu} &= 0,
\end{align}
which can be recast into the following forms:
\begin{align}
	u^\rho \nabla_\rho \beta &= \frac{n \rho_\mu - n_\mu (\rho+P)}{n_\mu \rho_\beta -n_\beta \rho_\mu }\nabla_\rho u^\rho, \\
	u^\rho \nabla_\rho (\beta \mu) &= \frac{-\beta n \rho_\beta + \mu n \rho_\mu + \beta n_\beta (\rho+P) - \mu n_\mu (\rho+P)}{\beta (n_\mu \rho_\beta - n_\beta \rho_\mu )} \nabla_\rho u^\rho, \\
	u^\rho \nabla_\rho u_\mu &= - \frac{P_\beta - \frac{\mu}{\beta} P_\mu}{\rho + P} \gamma_{\mu}{}^\nu \nabla_\nu \beta - \frac{P_\mu}{\beta(\rho + P)} \gamma_{\mu}{}^\nu \nabla_\nu (\beta \mu) - \frac{n}{\rho + P} u^\rho F_{\rho \mu}.
\end{align}
Now, $P_{(1)}$ and $j^\mu_{(1)}$ are written as
\begin{align}
	P_{(1)} & = {\bm \zeta} \nabla_\rho u^\rho, \\
	j^\mu_{(1)} & = {\bm \tau} \gamma^{\mu \nu} \nabla_\nu \beta - {\bm \chi}_1  \beta \gamma^{\mu \nu} \nabla_\nu  (\beta \mu) - {\bm \chi}_2 \beta^2  \gamma^{\mu \nu} u^\rho F_{\rho \nu}, \\
	 {\bm \zeta} & \equiv \tilde \zeta + \left( \frac{\partial P}{\partial \rho}\right)_n\tilde \lambda + \left( \frac{\partial P}{\partial n}\right)_\rho \tilde \nu, \label{zetabm} \\
	 {\bm \tau} & \equiv  \tilde \tau_1 - \tilde \tau_2, \\
	 {\bm \chi}_1 & \equiv  \tilde \chi_1 - \frac{n}{\beta(\rho+P)} \tilde \tau_2, \label{chibm} \\
	 {\bm \chi}_2 & \equiv  \tilde \chi_2 - \frac{n}{\beta(\rho+P)} \tilde \tau_2.
\end{align}
Here, we have used Eqs.~\eqref{prn} and \eqref{pnr}.
The $\bm \zeta$, $\bm \chi_i$ and $\bm \tau_i$ are hydrodynamical frame invariants by definition.

When imposing the KMS condition in the original hydrodynamical frame as presented in App.~\ref{KMSsummary}, one can further constrain the coefficients as
\begin{align}
	{\bm \tau} = 0, ~{\bm \chi}_1 = {\bm \chi}_2 \equiv {\bm \chi}.
\end{align}
Additionally, the unitarity constraints lead to 
\begin{align}
	{\bm \zeta}, {\bm \chi} \geq 0,
\end{align}
with 
\begin{align}
	x &= 1, \\
	y &= -\left( \frac{\partial P}{\partial \rho}\right)_n, \\
	z &= -\left( \frac{\partial P}{\partial n}\right)_\rho,
\end{align}
in Eq.~\eqref{positivdefeplam}, and
\begin{align}
	x &= - \frac{n}{\beta (\rho+P)}, ~y = 1,
\end{align}
in Eq.~\eqref{positivdefeplam2}.
Now, we arrive at the so-called Landau frame, where the first-order transport coefficients are characterized by three independent hydrodynamic parameters: the shear viscosity $\eta$, bulk viscosity $\zeta$, and charge conductivity $\chi$. 

In Ref.~\cite{Kovtun:2019hdm}, the ambiguity of the on-shell condition was emphasized. Specifically, it was shown that one can eliminate the term $\nabla_\mu u^\mu$ and express the constitutive relations in terms of either $u^\mu \nabla_\mu \beta$ or $u^\mu \nabla_\mu (\beta \mu)$. These correspond to the conditions $\tilde{\lambda} = \tilde{\zeta} = \tilde{\tau} = 0$ and $\tilde{\zeta} = \tilde{\nu} = \tilde{\tau} = 0$, respectively, thus modifying the hydrodynamic equations of motion. Therefore, on-shell solutions reintroduce ambiguities in the transport coefficients, which we refer to as on-shell transformations.

This freedom of on-shell transformation is distinct from that of a field redefinition, as it allows one to eliminate hydrodynamic frame invariants.

\subsection{Traditional hydrodynamical frames}

The derivation of the Landau frame presented above differs from the original prescription given by Landau and Lifshitz, which we now discuss.

Suppose one imposes the KMS condition after maximally eliminating coefficients via field redefinition. This corresponds to taking $T_{(1)}^{\mu\nu}u_\nu = j_{(1)}^{\mu}u_\mu = 0$, followed by the imposition of the second law of thermodynamics as outlined in the textbook by Landau and Lifshitz. The first condition leads to $\epsilon = \lambda_1 = \nu_1 = \nu_2 = \nu_3 = \nu_4 = \kappa_1 = \kappa_2 = \tau_3 = \tau_4 = 0$. The second condition yields $\lambda_2 = \lambda_1$, $\nu_5 = \nu_2$, $\kappa_i = \kappa$, $\tau_i = \tau$, and $\chi_i = \chi$. As a result, $T^{\mu\nu}$ and $J^\mu$ are expressed in terms of three non-vanishing transport coefficients: $\eta$, $\zeta$, and $\chi$. The noise coefficients vanish under the KMS condition except for $w_4 = \zeta$, $w_5 = \eta$, and $y_2 = \chi$. In other words, in the Landau frame, several hydrodynamical invariants are set to zero: $\tilde \lambda = \tilde \nu = \tilde \tau = 0$, while $\tilde \zeta = \zeta$ and $\tilde \chi = \chi$.

Similarly, the Eckart frame is realized by imposing $T_{(1)}^{\mu\nu}u_\nu u_\mu = j_{(1)}^{\mu} = 0$, which leads to $\epsilon = \lambda_1 = \nu_1 = \nu_2 = \nu_3 = \nu_4 = \tau_1 = \tau_2 = \chi_1 = \chi_2 = 0$. Then, under the KMS condition, the frame invariants are set to $\tilde \lambda = \tilde \nu = \tilde \chi = 0$, with $\tilde \zeta = \zeta$ and $\tilde \tau = -n\kappa/\beta(\rho + P)$.

These observations make it clear that the Landau and Eckart frames are not equivalent under hydrodynamical transformations, as they yield different hydrodynamical frame invariants. The two frames are related only up to a hydrodynamical frame transformation followed by an on-shell transformation.

At first glance, the freedom to solve lowest-order hydrodynamics arbitrarily might suggest that on-shell transformations are unphysical. However, such transformations alter the nature of the equations of motion and have implications for the analysis of stability and causality~\cite{Kovtun:2019hdm}. Indeed, earlier studies have concluded that traditional first-order hydrodynamics is neither stable nor causal, largely due to these ambiguities. In Sec.~\ref{sec:md}, we will demonstrate that the mode analysis remains unaffected by this freedom.

It is also important to note that neither hydrodynamical frame transformations nor on-shell transformations affect the noise sector. Therefore, once the KMS condition is imposed on the most general first-order action, the noise coefficients cannot be further reduced. This apparent inconsistency suggests that a full on-shell treatment of hydrodynamics—including noise fields—is necessary. We will return to this point in Sec.~\ref{gfsec}, where we show how the noise coefficients can indeed be further constrained.

\subsection{Ambiguity of rest frames}
\label{restframe}
In hydrodynamics, a “frame transformation” traditionally refers to the redefinition of $\beta^\mu$. In other contexts, however, the term “frame” typically denotes a coordinate system. In this work, we distinguish these usages by explicitly specifying “hydrodynamical” when referring to the former.

A coordinate transformation changes the field values evaluated at the same coordinate value, since the choice of the coordinate system at rest is not unique~\cite{Kodama:1984ziu}. In our setting, the generator of the coordinate transformation can be expanded as
\begin{align}
	\xi^\mu = \sum_{n=0} \xi^\mu_{(n)},
\end{align}
where $n$ indicates the order in the covariant derivative expansion. The change in field variables is then expressed through the Lie derivative associated with $\xi^\mu$. For example, the leading-order coordinate transformation is
\begin{align}
	x^\mu \to x^\mu + \pounds_\xi x^\mu = x^\mu + \xi^\mu.
\end{align}
Note that a coordinate system is represented by a set of scalar fields $x^\mu$. In our case, $\beta$ and $\mu$ are scalar fields, and $u_\mu$ is a one-form. Their transformations are given by
\begin{align}
	\delta \beta &= \pounds_\xi \beta = \xi^\rho \nabla_\rho \beta, \\
	\delta \mu &= \pounds_\xi \mu = \xi^\rho \nabla_\rho \mu, \\
	\delta u_\mu &= \pounds_\xi u_\mu = \xi^\rho \nabla_\rho u_\mu + u_\rho \nabla_\mu \xi^\rho.
\end{align}

At the lowest order in the derivative expansion, the transformation generator is given by
\begin{align}
	\xi_{(0)}^\mu = f(\beta, \mu) u^\mu.
\end{align}
The resulting “gauge transformation” generated by $\xi_{(0)}^\mu$ is then
\begin{align}
	\delta \beta &= f\, u^\rho \nabla_\rho \beta, \\
	\delta \mu &= -f\, u^\rho \mu \beta^{-1} \nabla_\rho \beta + f\, \beta^{-1} u^\rho \nabla_\rho (\beta \mu), \\
	\delta u_\sigma &= f\, u^\rho \nabla_\rho u_\sigma - (f_\beta - f_\mu \mu \beta^{-1}) \nabla_\sigma \beta - f_\mu \beta^{-1} \nabla_\sigma (\beta \mu),
\end{align}
which represents a subclass of field redefinitions.
In addition, a general diffeomorphism transformation of the metric is given by
\begin{align}
	g_{\mu \nu} \to g_{\mu\nu} + \nabla_\mu \xi_\nu + \nabla_\nu \xi_\mu.
\end{align}
Thus, a general coordinate transformation necessarily induces a perturbation of the metric.
Although gauge transformations of the metric are unavoidable, this point has not been emphasized in the literature.
Therefore, it is important to clearly distinguish general coordinate transformations from field redefinitions of thermodynamical variables.

\paragraph{Section summary.}
This section analyzed the twofold redundancy present in relativistic viscous hydrodynamics: the hydrodynamical frame transformation and the on-shell transformation. The former originates from field redefinitions of $\beta^\mu$ at first order, which leave the generating functional invariant while altering the form of the constitutive relations. In contrast, the on-shell transformation employs the leading-order equations of motion to further eliminate transport coefficients.  
Although this reduction does not change physical observables, the physical content must be expressed in terms of the frame-invariant combinations $\eta$, $\bm{\zeta}$, $\bm{\chi}_{1/2}$, and $\bm{\tau}_{1/2}$. After maximally reducing the coefficients using the KMS condition, the number of first-order transport coefficients is reduced to three: $\eta$, $\bm{\zeta}$, $\bm{\chi}$.
Nonetheless, it has been pointed out that on-shell transformations can change the character of the equations—from hyperbolic to elliptic—which is a potential source of instability. 
We also showed that, while these transformations do not affect the noise terms at the order considered, the full noise sector must be consistently solved on-shell to ensure compatibility with the KMS condition. Finally, we clarified the distinction between hydrodynamical and relativistic coordinate frames, showing that leading-order coordinate transformations constitute a subclass of hydrodynamical redefinitions, with a nontrivial transformation on the metric.

\section{Effective theory for hydrodynamical collective excitations}
\label{ngbaction}

In the previous sections, we constructed the generating functional for a general U(1) symmetric system and constrained its coefficients using the KMS symmetry and the unitarity of the UV theory.
We also discussed the redundancies associated with field redefinitions of $\beta^\mu$ and on-shell transformations.
In this section, we introduce the hydrodynamical mode and construct its effective quadratic action.

\medskip
A conventional and widely used approach to relativistic hydrodynamics treats $\beta$, $u^\mu$, and $\mu$ as thermodynamic variables, and analyzes their linear perturbations under the assumption of local thermal equilibrium.
This method has been highly successful in phenomenological applications, as it allows for straightforward analysis based on linearized conservation equations.

However, if one wishes to understand the underlying symmetry structure or systematically derive the dynamics of fluctuations, it is useful to revisit this formulation from the perspective of EFT.
In particular, the conventional hydrodynamic perturbations do not transform linearly under the symmetries of the generating functional, and thus the implementation of spacetime covariance—especially under Lorentz boosts—is not manifest.

In the presence of a finite temperature vector $\beta^\mu$, which defines a preferred frame, the full Lorentz symmetry is no longer manifest at the level of the background.
Nevertheless, the microscopic theory may still be Lorentz invariant, and the preferred frame should be understood as a feature of the ensemble.
From this viewpoint, the symmetry breaking is not explicit but rather spontaneous, as the thermal state selects a specific rest frame.

This motivates us to treat the hydrodynamical modes as nonlinear realizations of spontaneously broken spacetime symmetries, similar to Goldstone modes in other EFT contexts.
Indeed, sound waves in the fluid can be understood in this way, and their transformation under boosts acquires a nonlinear character.
This framework naturally leads to an effective action formulation, which respects the original symmetry structure at the level of the generating functional.

The construction of such an EFT has been initiated in Ref.~\cite{Crossley:2015evo}, which introduces the concept of emergent fluid spacetime to realize the full symmetry algebra, including fluctuation and dissipation.
While this approach is elegant and powerful, in this work we opt for a simpler formulation that focuses on symmetry realization in physical spacetime, in order to facilitate direct analysis of the hydrodynamical modes themselves.

\subsection{Symmetry structure in hydrodynamics}

The structure of SSB in hydrodynamics is more intricate than in the vacuum case due to the doubling principle inherent in the Schwinger--Keldysh formalism. Recent developments in the EFT perspective on hydrodynamics suggest that hydrodynamical modes are NG modes associated with so-called strong-to-weak spontaneous symmetry breaking (SWSSB)~\cite{Huang:2024rml}. Accordingly, we refer to $\pi^\mu$ as the NG mode in the context of SWSSB throughout this work.

The introduction of an NG mode for SSB of a global $U(1)$ scalar field theory proceeds as follows. We parameterize the complex scalar field as $\Phi = r e^{i\varphi}$. The global $U(1)$ symmetry is spontaneously broken by the vacuum. Excitations around a chosen vacuum phase $\varphi = \varphi_0$ are described by a field $\theta$, which nonlinearly realizes the symmetry broken by the vacuum via a constant shift $\theta \to \theta + \epsilon$. The radial mode $r$ is typically a heavy degree of freedom that is integrated out in hydrodynamics. Introducing $\theta$ leads to a shift in the external background field: $A_\mu \to A_\mu + \nabla_\mu \theta$. This transformation is distinct from a gauge transformation, since the $U(1)$ symmetry is global and $A_\mu$ is a nondynamical background field.

This structure is not specific to internal symmetry breaking. Consider a theory of a real scalar field $\phi$ in Minkowski spacetime. Suppose the dynamics admits a time-dependent solution $\phi = \bar \phi(t)$. Although the action is invariant under time translations, the solution breaks this symmetry spontaneously. The excitation around the background solution can then be described by an NG mode $\pi(x)$ via  
\begin{align}
	\phi(x) = \bar \phi(t + \pi(x)). \label{difemb}
\end{align}
Under time translations, the field $\pi(x)$ transforms with a constant shift, reflecting the nonlinearly realized time translation symmetry. These NG modes can be understood as reparametrizations of the original field variables, defined with respect to specific background solutions. More precisely, they correspond to the exponential map generated by the broken symmetry generators acting on the condensates that spontaneously break the symmetry.

The same idea can be applied to hydrodynamics. The local thermal vector $\beta^\mu$ is now reparametrized as
\begin{align}
	\beta^\mu = e^{-\pounds_\pi} \beta^\mu_c, \label{embed}
\end{align}
where $\beta^\mu_c = \beta_c \delta^\mu_0$ with a constant $\beta_c$, and $\pounds_\pi$ denotes the Lie derivative along $\pi^\mu$. Eq.~\eqref{embed} represents a local thermal state as a diffeomorphic embedding of a global thermal state. 

\medskip
For global symmetries, the NG modes have a clear physical meaning as excitations around the broken vacuum.
When gauging the symmetries, one can embed $\pi^\mu$ in the same way as in the global symmetry case; however, the physical meaning of the hydrodynamical modes depends on the choice of gauge.
In the conventional formulation~\cite{Crossley:2015evo}, the NG mode is eaten by the metric in the unitary gauge, where the fluid is unperturbed and the hydrodynamical theory is constructed.
The fluid is then defined by additional symmetry principles, with the thermal vector $\beta^\mu$ introduced as an emergent background quantity.

In contrast, in our approach we begin by specifying the physical thermal state through a given vector $\beta^\mu$, which encodes the local rest frame and temperature distribution.
We introduce $\beta^\mu$ in a gauge where the metric perturbations are eliminated, corresponding to the 't Hooft–Feynman gauge where the NG mode remains explicit.
The field $\pi^\mu$ is introduced directly into $\beta^\mu$ via Eq.~\eqref{embed}, and therefore represents physical fluctuations of the fixed thermal state.

Evaluating the Lie derivative,
\begin{align}
	\pounds_\pi \beta^\mu_c = \pi^\rho \partial_\rho \beta^\mu_c - \beta^\rho_c \partial_\rho \pi^\mu = - \beta_c \dot \pi^\mu,
\end{align}
we obtain
\begin{align}
	\beta^\mu = \beta_c (\delta^\mu_0 + \dot \pi^\mu).
\end{align}
Under a general coordinate transformation $x^\mu \to x^\mu- \xi^\mu(x)$, we have
\begin{align}
	\beta^\mu(x) \to  \beta_c \delta^\mu_0  + \beta_c(\dot \pi^\mu + \dot \xi^\mu) +\mathcal O(\xi^2).
\end{align}
Thus, general coordinate transformations are nonlinearly realized on $\pi^\mu$:
\begin{align}
	\pi'^\mu(x) = \pi^\mu(x) + \xi^\mu (x).\label{def:nontl}
\end{align}

The local temperature, velocity, and chemical potential are defined as
\begin{align}
	\beta &\equiv \sqrt{-g_{\mu\nu}\beta^\mu \beta^\nu},
	\\
	u^\mu &\equiv \frac{\beta^\mu}{\beta},
	\\
	\mu &\equiv \frac{A_\mu \beta^\mu}{\beta}.
\end{align}
For simplicity, we take the background fields as
\begin{align}
	g_{\mu\nu} = \eta_{\mu\nu},~A_\mu = \mu_c \delta^0_\mu.
\end{align}
In this case, we find
\begin{align}
	\beta &= \beta_c (1+\dot \pi^0),
	\\
	u^\mu &= (1, \dot \pi^i),
	\\
	\mu &= \mu_c + \dot \theta.
\end{align}
Thus, hydrodynamical perturbation variables are expressed by the NG modes, and now the covariance becomes manifest.
While $\mu$, $\beta$, and $u^\mu$ are zeroth-order quantities in the constitutive relations, they are first order in derivatives of the NG modes.

The induced metric $\gamma^{\mu\nu}$, defined with respect to $u^\mu$, is distinct from the spatial metric on constant time hypersurfaces.
We find
\begin{align}
	\gamma^{00} 	= 0 , ~ \gamma^{0i} 	=  \dot \pi^i ,~\gamma^{ij} 	=  \delta^{ij} .
\end{align}

Hydrodynamical coefficients are perturbed via $\beta$ and $\mu$.
For example, the energy density $\rho$ expands as
\begin{align}
	\rho(\beta,\mu) = \rho_{\rm c}   
	+
	 \beta_{\rm c} \rho_{\beta} \dot \pi^0  +
	\rho_{\mu} \dot \theta,
\end{align}
where $X_\mu \equiv (\partial X/\partial \mu)_\beta$ and $X_\beta \equiv (\partial X/\partial \beta)_\mu$ for $X=P,\rho,n$. 
Hereafter, for notational simplicity, we omit the subscript “c” and redefine the background values as $\beta\equiv \beta_{\rm c}$ and $\mu \equiv \mu_{\rm c}$.

\medskip
So far, we have only discussed the SSB of the retarded sector in the Schwinger-Keldysh formalism. 
This sector has advanced counterparts, often referred to as “noise” fields.
The Ward identities for the retarded fields,
\begin{align}
	 \nabla_\mu J^\mu &= 0,
	\\
	\nabla_\mu T^{\mu\nu} + J_\mu F^{\mu\nu} &=0,
\end{align}
imply a symmetry of the generating functional under the following transformations:
\begin{align}
	\delta_{\bar \xi,\bar \Lambda}  g_{\mu\nu} & = 0 ,\label{transf:g}
	\\
	\delta_{\bar \xi,\bar \Lambda} \bar g_{\mu\nu} & = \nabla_\mu \bar \xi_\nu + \nabla_\nu \bar \xi_\mu ,\label{transf:g}
	\\
	\delta_{\bar \xi,\bar \Lambda} A_{\mu} &= 0 ,
	\\
	\delta_{\bar \xi,\bar \Lambda} \bar A_{\mu} &= -\nabla_\mu \bar \Lambda + \bar \xi^\nu \nabla_\nu A_{\mu} + A_\nu \nabla_\mu \bar \xi^{\nu}. \label{transf:A}
\end{align}
These transformations reflect redundancies in the doubled degrees of freedom on the CTP.
In open systems, this symmetry is generally broken and the energy momentum conservation is not ensured anymore.

One can restore the above symmetries by introducing Stueckelberg fields:
\begin{align}
	\bar G_{\mu\nu} &= \bar g_{\mu\nu}  - \nabla_\mu {\bar \pi}_\nu - \nabla_\nu {\bar \pi}_\mu,
	\\
	\bar C_\mu &= \bar A_\mu + \nabla_\mu \bar \theta - {\bar \pi}^\nu \nabla_\nu A_{\mu} - A_\nu \nabla_\mu {\bar \pi}^{\nu}.
\end{align}
Since SSB occurs in the retarded sector, these Stueckelberg fields can be interpreted as the advanced-sector counterparts of the NG modes.

A more refined analysis is possible by carefully treating the duplicated fields defined on the CTP.
To summarize, the set of variables $\Phi = \{\beta, u^\mu, \mu, F_{\mu\nu}, g_{\mu\nu} \}$ and $\bar \Phi = \{\bar G_{\mu\nu}, \bar C_\mu \}$, along with their covariant derivatives, constitute the symmetry-preserving building blocks for the dynamical action of NG modes.

\subsection{Effective action for Nambu-Goldstone modes}
\label{effac}

When NG modes are set to zero, the effective action must reproduce the UV generating functional~\eqref{locaction}.
Hence, the effective action for the NG modes should be written as~\cite{Glorioso:2017fpd}
\begin{align}
	I[\Phi,\bar \Phi ]  = \int d^4 x \sqrt{-\det g}\sum_{n=1} \mathcal L_n,\label{pathI}
\end{align}
where we defined
\begin{align}
	\mathcal L_1 &= \frac{1}{2} T^{\mu \nu}[\Phi]\bar G_{\mu \nu} + J^\mu [\Phi] \bar C_\mu,
	\\
	\mathcal L_2 &=  \frac{i}{4} W^{\mu \nu \rho \sigma}[\Phi]\bar G_{\mu \nu}\bar G_{\rho \sigma} + i X^{\mu \nu \rho }[\Phi]\bar G_{\mu \nu}\bar C_\rho+ i Y^{\mu  \rho }[\Phi]\bar C_{\mu }\bar C_\rho.
\end{align}
By construction, one finds $T$ and $J$ by taking the variation of $e^{iI}$ with respect to $\bar g$ and $\bar A$.
$W$, $X$ and $Y$ are the correlation functions of $T$ and $J$.

By using these expressions in the NG modes, we expand the effective action in the following form:
\begin{align}
	I = \sum_{nml}I_d^{(n,m,l)}+I_{s}^{(n,m,l)}+I_{ds}^{(n,m,l)},
\end{align}
where $n$, $m$, and $l$ count the number of the noise fields, covariant derivatives, and the physical NG modes.
The subscripts $d$ and $s$ imply the noise dynamical variables and noise external sources.
$ds$ means the cross term of $s$ and $d$, which we do not consider in the present semi-classical order.
In the following, we closely look at the effective action order by order in the NG modes and source.

\paragraph{Source terms---.}

Let us start from the leading order effective action giving the perfect fluid component:
\begin{align}
I_s^{(1,0,0)}=\frac{1}{2} \int d^4 x  \Bigg[  \rho  \bar g_{00}
 +   P \delta^{ij}\bar g_{ij} +2n \bar A_0
 \Bigg].	
\end{align}
When taking the variation with respect to the noise metric and gauge field, the leading-order energy density, pressure, and number density are found.
Similarly, the variation in $\bar A_0$ yields $n$.
Next, let us consider the NLO, including physical NG modes.
Firstly, we find
\begin{align}
I^{(1,0,1)}_{s} = \int d^4 x & \Bigg[	
  \left( \frac{ \beta \rho_\beta  }{2 }\dot\pi^0 + \frac{\rho_\mu}{2}\dot \theta \right)\bar g_{00}
 +  (\rho + P) \bar g_{0i} \dot\pi^i
 +
 \left(\frac{\beta P_\beta }{2 }\dot\pi^0 + \frac{ P_\mu }{2 }\dot\theta \right)\delta^{ij}\bar g_{ij}
 \notag 
 \\
 & \bar A_0(n_\mu \dot \theta  +\beta n_\beta \dot \pi^0 ) + \bar A_i(n \dot \pi^i ) \Bigg].\label{sous1}
\end{align}
The variation with respect to the noise metric gives the lowest-order density and pressure fluctuations and the 3-velocity written by the NG modes.
Similarly, we get the effective action in the first-order gradient expansion:

\begin{align}
I^{(1,1,1)}_{s}&=\int d^4 x \Bigg[	
  \bar g_{00} \left(
  \frac{1}{2} \left(\epsilon  \ddot\pi^0
 - \lambda_1  \partial_i \dot\pi^i\right)
 - \frac{1}{2} \beta  \nu_4 \left(\ddot\theta
 + \mu  \ddot\pi^0\right)
 \right)
 \notag 
 \\
 &
 +\bar g_{0i} \left(
  \kappa_1  \partial^i \dot\pi^0
 - \kappa_2 \ddot\pi^i
 - \beta  \tau_3  \left(\partial^i \dot\theta
 + \mu  \partial^i \dot\pi^0\right)
 \right)
 \notag 
 \\
 &
 - \eta  \bar g_{ij} \partial^j\dot\pi^i
 + \delta^{ij}\bar g_{ij} \left(
  \frac{1}{6} \left(3 \lambda_2  \ddot\pi^0
 + (
 - 3 \zeta 
 + 2 \eta ) \partial_k\dot\pi^k\right)
 - \frac{1}{2} \beta  \nu_5 \left(\ddot\theta
 + \mu  \ddot\pi^0\right)
 \right)
 \notag 
 \\
 &
 + \bar A_0 \beta \left(\nu_1 \ddot\pi^0
 - \beta  \nu_3 \left(\ddot\theta
 + \mu  \ddot\pi^0\right)
 - \nu_2 \partial_i \dot\pi^i\right)
 \notag 
 \\
 &
 + \bar A_i \beta \left(
 - \beta  \chi_1  \left(\partial^i \dot\theta
 + \mu  \partial^i \dot\pi^0\right)
 + \tau_1  \partial^i \dot\pi^0
 - \tau_2 \ddot\pi^i\right)
 \Bigg].\label{eq:338}
\end{align}
By differentiating Eq.~\eqref{eq:338}, one finds the first order fluctuations to $T$ and $J$.
The noise term is already presented in Eq.~\eqref{noiseactiong}.

\paragraph{Action for the NG modes---.}

The lowest order effective action for ${\bar \pi}$ and $\bar \theta$ is
\begin{align}
I_d^{(1,0,0)}=- \int d^4 x  \Bigg[  (\rho - \mu n)  \dot{ {\bar \pi}}_{0}
 +  P \partial^i{\bar \pi}_i
 - n \dot{ \bar \theta}
\Bigg].	
\end{align}
By integrating by parts, we find
\begin{align}
I^{(1,0,0)}_d=\int d^4 x   
  \Bigg[ (\dot\rho - \mu \dot n){\bar \pi}_{0} + \dot n\bar \theta	\Bigg].
\end{align}
Note that we redefined $\mu$ as $\mu\equiv \mu_{\rm c}$, which is now constant.
The variations with respect to ${\bar \pi}$ and $\bar \theta$ yield the energy and number conservations in flat spacetime:
\begin{align}
	\dot\rho = 0,~\dot n = 0.\label{conti}
\end{align}
The lowest-order quadratic action for the NG modes is
\begin{align}
I^{(1,0,1)}_{d} &= \int d^4 x \Bigg[	
 \dot{\bar \pi}_0 \left(\dot\theta \left(\mu  n_\mu
 - \rho _\mu\right)
 + \beta  \dot\pi^0 \left(\mu  n_\beta
 - \rho_\beta\right)\right)
 - \partial_i\bar \pi_0 \dot\pi^i (\rho+P-\mu  n)
 \notag 
 \\
 &
   - \dot{\bar \pi}_i \dot\pi^i (\rho + P)
 + \partial^i{\bar \pi}_i \left(
 - \dot\theta P_\mu
 - \beta  \dot\pi^0 P_\beta\right)
 \notag 
 \\
 &
 +n \partial_i\bar \theta \dot\pi^i
 + \dot{\bar \theta} \left(\dot\theta n_\mu
 + \beta  \dot\pi^0 n_\beta \right)
 \Bigg].
\end{align}
This action describes the NG modes in the perfect fluid.
The NLO in the gradient is
\begin{align}
I^{(1,1,1)}_{d}&=\int d^4 x \Bigg[
  \dot{{\bar \pi}}_0 \left(
 - \epsilon  \ddot\pi^0
 + \lambda_1  \partial_i\dot\pi^i
 + \nu_1 \beta \mu  \ddot\pi^0
 + \beta \nu_4 \left(  \ddot\theta
 +  \mu  \ddot\pi^0\right)
 - \beta \mu  \nu_2 \partial_i\dot\pi^i
 - \beta^2  \mu  \nu_3 \left(\ddot\theta
 + \mu  \ddot\pi^0\right)
 \right)
 \notag 
 \\
 &
 + \partial_i{\bar \pi}_0 \left(
 -\kappa_1 \partial^i\dot\pi^0
 +\kappa_2 \ddot\pi^i
 +
 \tau_1 \beta  \mu  \partial^i\dot\pi^0 
 + \tau_3 \beta \left(\partial^i\dot\theta
 +  \mu  \partial^i\dot\pi^0
 \right)
 -\tau_2  \beta \mu  \ddot\pi^i
 - \beta^2  \mu  \chi_1  \left(\partial^i\dot\theta
 + \mu  \partial^i\dot\pi^0\right)
 \right)
  \notag 
 \\
 &
+
  \dot{{\bar \pi}}_i \left(  
 - \kappa_1 \partial^i\dot\pi^0
 + \kappa_2 \ddot\pi^i
 +
 \beta  \tau_3  \left(\partial^i\dot\theta
 + \mu  \partial^i\dot\pi^0\right)
 \right)
  \notag 
 \\
 &
 + \eta  \partial_j{\bar \pi}_i \left(\partial^j\dot\pi^i
 + \partial^i\dot\pi^j\right)
 + \partial^i{\bar \pi}_i \left(
 - \lambda_2  \ddot\pi^0
 + \frac{1}{3} (3 \zeta 
 - 2 \eta ) \partial_j\dot\pi^j
 + \beta  \nu_5 \left(\ddot\theta
 + \mu  \ddot\pi^0\right)
 \right)
 \notag 
 \\
 &
 + \dot{\bar \theta} \beta \left(\nu_1 \ddot\pi^0
 - \beta  \nu_3 \left(\ddot\theta
 + \mu  \ddot\pi^0\right)
 - \nu_2 \partial_i\dot\pi^i\right)
 \notag 
 \\
 &+\partial_i\bar \theta \beta\left(
 - \beta  \chi_1  \left(\partial^i\dot\theta
 + \mu  \partial^i\dot\pi^0\right)
 + \tau_1  \partial^i\dot\pi^0
 - \tau_2 \ddot\pi^i\right).
\end{align}
The first-order action adds diffusion effects.
Note that this action is first order in $\varepsilon$ but is third order with respect to the derivative on the NG modes.

Expanding the effective action to second order in $\bar \pi_{\mu}$, we find the noise action for the NG modes:
\begin{align}
 I^{(2,0,0)}_{d} &= \int d^4 x	\frac{i}{  \beta}\Bigg[
	 \epsilon \dot{ {\bar \pi}}_{0}{}^2
	 + \zeta (\partial^i{\bar \pi}_i )^2
  + 2  \lambda \dot{ {\bar \pi}}_{0} \partial^i{\bar \pi}_i
   +  \kappa  \left(\partial_i{\bar \pi}_{0}+\dot{{\bar \pi}}_i\right)^2
\notag
\\
&
 +  \eta \left( \left(\partial_j{\bar \pi}_i \right)^2
 +    \partial^j{\bar \pi}_i \partial^i\bar \pi_{j}
 - \frac{2}{3}   \partial^i{\bar \pi}_i \partial^j\bar \pi_{j}\right)
 \notag 
 \\
 &
 +
 \nu_3 (\dot{\bar \theta}+\mu \dot {\bar \pi}_0)^2
 +
 \chi(\partial_i\bar \theta + \mu \partial_i \bar \pi_0)^2
 \notag 
 \\
 &
 -
 2\nu_1 \dot{\bar \pi}_0(\dot{\bar \theta}+\mu \dot {\bar \pi}_0)
 -
 2\tau
 \left(\partial_i{\bar \pi}_{0}+\dot{{\bar \pi}}_i\right)(\partial^i\bar \theta + \mu \partial^i \bar \pi_0)
 -
 2\nu_2 (\dot{\bar \theta}+\mu \dot {\bar \pi}_0)\partial^i \bar \pi_i
\Bigg].
\end{align}
The noise action for the NG modes is also non negative definite, as expected from the construction.

Before going to the mode analysis, let us explain the present expansion scheme.
In the first-order hydrodynamical theory, there are two energy scales: $M_{\rm SSB}$ and $M_{\rm diss.}$.

$M_{\rm SSB}$ is the energy of spontaneous symmetry breaking, typically $\rho+P \sim M_{\rm SSB}^4$, which is the decay constant in nonlinear realization theory.
For a conformal fluid, $M_{\rm SSB} \sim \beta^{-1}$.
The perturbative description is not applicable for $\omega > M_{\rm SSB}$, which is typically the energy scale of a thermal fast mode.
In fact, one can canonically normalize the NG mode as $\pi_{\rm c} \sim \sqrt{\rho+P}\pi$, and the action is expanded by $(\partial \pi_{\rm c}/M^2_{\rm SSB})^n$.
Hence, the perturbative expansion is justified for $\omega \ll M_{\rm SSB}$.
$\eta\sim M_{\rm diss.}^3$ is the energy scale of dissipation determined by the mean-free-path~(MFP) of the fast modes.
The order counting parameter of the gradient expansion is $\varepsilon \sim \omega M_{\rm diss.}^3/M_{\rm SSB}^4$, and the gradient expansion is valid in the regime where $\varepsilon \ll 1$.

\paragraph{Section summary.}
In this section, we argued that the spontaneous breaking of spacetime symmetry occurs in the presence of a local thermal state characterized by the local four-vector $\beta^\mu$.
We then demonstrated that perturbations in $\beta^\mu$ can be expressed in terms of NG modes, analogous to the embedding of NG modes in global U(1) theories.
The spacetime NG mode $\pi^\mu$ nonlinearly realizes diffeomorphism symmetry, enabling a manifestly covariant formulation of hydrodynamical perturbations.
We also introduced the noise counterparts of the NG modes and constructed the corresponding hydrodynamical effective action.
The remaining task is a straightforward, albeit technically involved, mode analyis of this effective action.

\section{Mode analysis}
\label{sec:md}

In the previous section, we obtained the most general quadratic action for the first-order U(1) hydrodynamical collective modes.
Let us consider the mode analysis and discuss the stability of thermal states.

\subsection{Action in Fourier space}

In this paper, the Fourier integral of the NG modes is defined as
\begin{align}
	\pi^M(x) = \int \frac{d\omega}{2\pi} \int \frac{d^3 k}{(2\pi)^3}e^{-i\omega t + i\mathbf k\cdot \mathbf x} \pi^M_{\omega,\mathbf k},
\end{align}
with $M=\mu,4$ and we assigned $M=4$ for $\theta$.
In the plane wave mode analysis, ${\rm Im}~\omega > 0$ implies the exponential growth of perturbations; therefore, the criteria of stability is given by ${\rm Im}~\omega \leq 0$. 
Hereafter, we consider Fourier space, and the arguments for the NG modes are suppressed for notational simplicity.
Then, the dynamical action is recast into
\begin{align}
I^{(1,0,1)}_{d} + I^{(1,1,1)}_{d} &= \int \frac{d\omega}{2\pi}  \int \frac{d^3 k}{(2\pi)^3}	 
 ({\bar \pi}_{0}^*~{\bar \pi}_i^*~\bar \theta^*) 
2\mathcal E 
\begin{pmatrix}
	\pi^{0} \\
	 \pi^{j} \\
	  \theta
\end{pmatrix}
\label{sec:action},
\end{align}
where we defined
\begin{align}
2\mathcal{E}^{0}{}_0&=\beta  \omega ^2 (\mu  n_\beta-\rho_\beta )+ i \omega ^3 \left(\beta ^2 \mu ^2 \nu_{3}-  \beta  \mu  (\nu_{1} + \nu_4)+ \epsilon \right)\notag 
\\
&+ik^2 \omega  \left( \beta ^2 \mu ^2 \chi -\beta  \mu  (\tau_1+\tau_3) + \kappa \right) 
,
\\
2\mathcal{E}^{0}{}_i&= k_i \omega  (\rho + P - \mu n)+ i k_i \omega ^2 (\kappa_2 + \lambda_1  - \beta \mu  (  \nu_{2}+   \tau_2 ))
,
\\
2\mathcal{E}^{0}{}_4&= \omega ^2 (\mu  n_\mu-\rho_\mu ) + i \beta \omega ^3 \left( \beta \mu  \nu_{3}-  \nu_{4}\right)+i \beta k^2 \omega  \left(  \beta \mu  \chi_1 - \tau_3  \right)
,
\\
2\mathcal{E}^{i}{}_0&= \beta  \omega k_i P_\beta - i k_i \omega ^2 ( \kappa_1 +\lambda_2 -\beta  \mu  (\nu_{5}+\tau_3 ) ) 
,
\\
2\mathcal{E}^{i}{}_j&=  - \left(\omega ^2 (\rho +\rho )+i \kappa  \omega ^3+i \eta  k^2 \omega  
	\right)\delta^i{}_j
	-\frac{1}{3} i k_i k_j \omega  (3 \zeta +\eta ),
	,
\\
2\mathcal{E}^{i}{}_4&=k_i P_\mu  \omega +i \beta  k_i \omega ^2 (\nu_{5}+\tau_3 )
,
\\
2\mathcal{E}^{4}{}_0&=i\omega ^3 \left(\beta ^2 \mu  \nu_{3}- \beta  \nu_{1}\right)+i\beta \omega k^2 \left(\beta   \mu  \chi_1 -   \tau_1 \right)+\beta  n_\beta \omega ^2
,
\\
2\mathcal{E}^{4}{}_i&=-\omega k_i n  -i \beta  \omega ^2 k_i  (\nu_{2}+\tau_2 )
,
\\
2\mathcal{E}^{4}{}_4&=n_\mu \omega ^2 + i \beta ^2 \omega ^3 \nu_{3} +i \beta ^2 \omega k^2 \chi_1.  
\end{align}
We are interested in thermal states, so the KMS condition is assumed.
However, we explicitly distinguish the equivalent hydrodynamical coefficients, such as $\lambda_1$ and $\lambda_2$, to discuss field redefinition later.
The thermodynamical relations \eqref{thr} and integrability condition~\eqref{ibc} yield
\begin{align}
	\mathcal E^0{}_i = -\mathcal E^i{}_0,~\mathcal E^0{}_4 = \mathcal E^4{}_0,~\mathcal E^i{}_4 = -\mathcal E^4{}_i.
\end{align}
Similarly, the leading order noise action is
\begin{align}
I^{(2,0,0)}_{d} &= \int \frac{d\omega}{2\pi}  \int \frac{d^3 k}{(2\pi)^3}	 
({\bar \pi}_{0}^*~{\bar \pi}_i^*~{\bar \theta}^*) 
\mathcal N 
\begin{pmatrix}
	\bar \pi_{0} \\ \bar \pi_{j} \\ \bar \theta
\end{pmatrix},
\end{align}
where we have introduced the noise matrix
\begin{align}
\mathcal N^{00} &= \frac{i}{\beta}\left(\omega^2(\mu ^2 \nu_3 -2 \mu  \nu_1 +\epsilon) +(\kappa +\mu ^2 \chi -2 \mu  \tau) k^2\right)
,
\\
\mathcal N^{0i} &= -\frac{i}{\beta}\omega k^i(\lambda+\kappa - \mu \tau - \mu \nu_2)
,
\\
\mathcal N^{04} &= \frac{i}{\beta}( k^2(-\tau+\mu\chi) +\omega^2 (-\nu_1 +\mu \nu_3) ) 
,
\\
\mathcal N^{ij} &= \frac{i}{\beta}\left[ (\kappa \omega^2 
 +  \eta k^2)\delta^{ij} 
 + \left(\zeta + \frac{1}{3} \eta\right) k^ik^j\right]
 ,
\\
\mathcal N^{i4} &= \frac{i}{\beta}\omega k^i \left[ \tau + \nu_2 \right]
 ,
 \\
\mathcal N^{44} &= \frac{i}{\beta}\left[ \omega^2 \nu_3 + k^2 \chi \right]
.
\end{align}
Our effective action is found to the noise order, which is a new result of this work.

\subsection{Constraint equations}

In the traditional analysis, one often computes the determinant of $\mathcal E$, which is recast into 
\begin{align}
	\det ( \mathcal E ) \propto \det ( \mathcal E^{\rm long.} ) \det ( \mathcal E^{\rm trans.} ), 
\end{align}
where the right hand side implies the product of the determinants of the longitudinal and transverse modes.
The dispersion relations are found from $\det ( \mathcal E )=0$.
The redundancy in the variables suggests the necessity of solving the constraint equations.
In this section, we carefully solve the constraint equations given by the continuity equation and change conservation.

In the action formalism, the dynamical and noise actions are written as
\begin{align}
	I_d  = &\int \frac{d\omega}{2\pi} \int \frac{d^3k}{(2\pi)^3}
	\bm{\pi}^\dagger G^{-1} \bm{\pi}	,
\end{align}
where we introduced 
\begin{align}
	\bm {\pi} = \begin{pmatrix}
		\pi^\mu
		\\
		\theta
		\\
		\bar \pi_{\mu}
		\\
		\bar \theta
	\end{pmatrix},
~
	G^{-1} = 
	\begin{pmatrix}
		0 & \mathcal E^* 
		\\
		\mathcal E &\mathcal N 
	\end{pmatrix}.
\end{align}
We want to find $G$, but $G^{-1}$ is not invertible as it contains the redundant variables.
Hence, we need to reduce the variables by solving the constraint equations
\begin{align}
	\frac{\delta I_d }{\delta \pi^{0*}} &={\bar \pi}_{0}(\mathcal E^0{}_0)^* + {\bar \pi}_i(\mathcal E^i{}_0)^* + \bar \theta (\mathcal E^{4}{}_0)^*=0
	,
	\\
		\frac{\delta I_d }{\delta \theta^*} &={\bar \pi}_{0}(\mathcal E^0{}_4)^* + {\bar \pi}_i(\mathcal E^i{}_4)^* + \bar \theta (\mathcal E^{4}{}_4)^*=0
	,
	\\
	\frac{\delta I_d }{\delta \bar \pi^*_{0}} &= 2\mathcal E^0{}_0\pi^0 +2 \mathcal E^0{}_i\pi^i+ 2\mathcal E^0{}_4\theta + \mathcal N^{00}{\bar \pi}_{0}+\mathcal  N^{0i}{\bar \pi}_i+\mathcal  N^{04}{\bar \theta} = 0,\label{const15}
		\\
	\frac{\delta I_d }{\delta \bar \theta^*} &= 2\mathcal E^4{}_0\pi^0 +2 \mathcal E^4{}_i\pi^i+ 2\mathcal E^4{}_4\theta +\mathcal  N^{40}{\bar \pi}_{0}+\mathcal  N^{4i}{\bar \pi}_i+\mathcal  N^{44}{\bar \theta} = 0,
	\label{const2}
\end{align}
where, as defined above, the 4th index corresponds to $\theta$.
Note that the complex conjugates are not regarded as independent variables.
For classical dynamics, the noise field is set to 0 so that the first two equations are trivial.
Then the third and fourth equations reduce to the linearized continuity equation and number conservation.
In the present case, we want to eliminate the redundant variables up to the noise order consistently, so we need to account for the noise correction in Eqs.~\eqref{const15} and~\eqref{const2}.
These equations are solved as
\begin{align}
	\pi^0 &= L^{0}{}_i\pi^i + M^{0i} \bar \pi_i,
	\\
	\theta &= L^4{}_i\pi^i  + M^{4i} \bar \pi_i,
	\\
	\bar \pi_0 & = \bar \pi_i L^{i}{}_0,
	\\
	\bar \theta & =\bar \pi_i L^{i}{}_4,
\end{align}
where we defined
\begin{align}
L^{0}{}_i &=-  \frac{ \mathcal E^0{}_i \mathcal E^4{}_4-\mathcal E^0{}_4 \mathcal E^4{}_i}{\mathcal E^0{}_0 \mathcal E^4{}_4- \mathcal E^0{}_4 \mathcal E^4{}_0},
\\
L^{4}{}_i &= \frac{ \mathcal E^0{}_i \mathcal E^4{}_0-\mathcal E^0{}_0 \mathcal E^4{}_i}{\mathcal E^0{}_0 \mathcal E^4{}_4-\mathcal E^0{}_4 \mathcal E^4{}_0},
\\
L^{i}{}_0 &= - (L^0{}_i)^*,
\\
L^{i}{}_4 &= - (L^4{}_i)^*,
\\
2M^{0i}	&=-\frac{\mathcal N^{4i} \mathcal E^0{}_4-\mathcal N^{0i} \mathcal E^4{}_4}{(\mathcal E^0{}_4)^2-\mathcal E^0{}_0 \mathcal E^4{}_4}
	\notag 
	\\
	&
	+\frac{\mathcal N^{44} \mathcal E^0{}_4 ((\mathcal E^0{}_0)^* (\mathcal E^4{}_i)^*-(\mathcal E^0{}_4)^* (\mathcal E^0{}_i)^*)}{\left((\mathcal E^0{}_4)^2-\mathcal E^0{}_0 \mathcal E^4{}_4\right) \left((\mathcal E^0{}_4)^{*2}-(\mathcal E^0{}_0)^* (\mathcal E^4{}_4)^*\right)}
	\notag 
	\\
	&+\frac{\mathcal N^{00} ((\mathcal E^0{}_4)^* \mathcal E^4{}_4 (\mathcal E^4{}_i)^*-(\mathcal E^0{}_i)^* \mathcal E^4{}_4 (\mathcal E^4{}_4)^*)}{\left((\mathcal E^0{}_4)^2-\mathcal E^0{}_0 \mathcal E^4{}_4\right) \left((\mathcal E^0{}_4)^{*2}-(\mathcal E^0{}_0)^* (\mathcal E^4{}_4)^*\right)}
	\notag 
	\\
	&+\frac{\mathcal N^{04} (-(\mathcal E^0{}_0)^* \mathcal E^4{}_4 (\mathcal E^4{}_i)^*-\mathcal E^0{}_4 (\mathcal E^0{}_4)^* (\mathcal E^4{}_i)^*+\mathcal E^0{}_4 (\mathcal E^0{}_i)^* (\mathcal E^4{}_4)^*+(\mathcal E^0{}_4)^* (\mathcal E^0{}_i)^* \mathcal E^4{}_4)}{\left((\mathcal E^0{}_4)^2-\mathcal E^0{}_0 \mathcal E^4{}_4\right) \left((\mathcal E^0{}_4)^{*2}-(\mathcal E^0{}_0)^* (\mathcal E^4{}_4)^*\right)},
\\
2M^{4i}&=
\frac{\mathcal N^{4i} \mathcal E^0{}_0-\mathcal N^{0i} \mathcal E^0{}_4}{(\mathcal E^0{}_4)^2-\mathcal E^0{}_0 \mathcal E^4{}_4}
\notag 
\\
&
+\frac{\mathcal N^{44} \mathcal E^0{}_0 ((\mathcal E^0{}_4)^* (\mathcal E^0{}_i)^*-(\mathcal E^0{}_0)^* (\mathcal E^4{}_i)^*)}{\left((\mathcal E^0{}_4)^2-\mathcal E^0{}_0 \mathcal E^4{}_4\right) \left((\mathcal E^0{}_4)^{*2}-(\mathcal E^0{}_0)^* (\mathcal E^4{}_4)^*\right)}
\notag 
\\
&
+\frac{\mathcal N^{00} \mathcal E^0{}_4 ((\mathcal E^0{}_i)^* (\mathcal E^4{}_4)^*-(\mathcal E^0{}_4)^* (\mathcal E^4{}_i)^*)}{\left((\mathcal E^0{}_4)^2-\mathcal E^0{}_0 \mathcal E^4{}_4\right) \left((\mathcal E^0{}_4)^{*2}-(\mathcal E^0{}_0)^* (\mathcal E^4{}_4)^*\right)}
\notag 
\\
&+\frac{\mathcal N^{04} (\mathcal E^0{}_0 (\mathcal E^0{}_4)^* (\mathcal E^4{}_i)^*-\mathcal E^0{}_0 (\mathcal E^0{}_i)^* (\mathcal E^4{}_4)^*+(\mathcal E^0{}_0)^* \mathcal E^0{}_4 (\mathcal E^4{}_i)^*-\mathcal E^0{}_4 (\mathcal E^0{}_4)^* (\mathcal E^0{}_i)^*)}{\left((\mathcal E^0{}_4)^2-\mathcal E^0{}_0 \mathcal E^4{}_4\right) \left((\mathcal E^0{}_4)^{*2}-(\mathcal E^0{}_0)^* (\mathcal E^4{}_4)^*\right)}.
\end{align}
By eliminating $\pi^0$, ${\bar \pi}_{0}$, $\theta$ and $\bar \theta$, the quadratic action is recast into
\begin{align}
	\bm{\pi}^\dagger G^{-1} \bm{\pi}	 =& \frac{\delta I_d }{\delta \pi^{0}}\pi^{0} + \frac{\delta I_d }{\delta \theta}\theta
	+2{\bar \pi}_i^*\mathcal E^i{}_j\pi^j
	+2{\bar \pi}_{0}^*\mathcal E^0{}_j\pi^j
	+2{\bar \theta}^*\mathcal E^4{}_j\pi^j
	\notag 
	\\
	&+
	{\bar \pi}_i^*  \mathcal N^{ij} \bar \pi_{j} 
	+ {\bar \pi}_i^* \mathcal N^{iN} {\bar \pi}_{N}
	+ {\bar \pi}_{M}^* \mathcal N^{Mj} {\bar \pi}_j
	+ {\bar \pi}_{M}^* \mathcal  N^{MN} {\bar \pi}_{N}
	\notag 
	\\
	=&
2{\bar \pi}_i^*\tilde{\mathcal E}^i{}_j \pi^j
	+
	{\bar \pi}_i^*  \tilde{\mathcal N}^{ij} \bar \pi_{j},
\end{align}
where $N, M=0,4$, and the reduced matrixes are defined as
\begin{align}
	\tilde{\mathcal E}^i{}_j&= \mathcal E^i{}_j + (L^{i}{}_M)^* \mathcal E^M{}_j ,
	\\
	\tilde {\mathcal N}^{ij}&=\mathcal N^{ij} 	+  \mathcal N^{iN} L^{j}{}_N
	+ (L^{i}{}_M)^* \mathcal N^{Mj} 
	+ (L^{i}{}_M)^* \mathcal  N^{MN} L^{j}{}_N
.\label{deftildeN}
\end{align}
The effective action is further simplified by separating the longitudinal and transverse modes.
Consider the scalar–vector decomposition in flat space~\cite{Kodama:1984ziu}
\begin{align}
	{\bar \pi}_i = ik_i {\bar \pi}^{\rm L} +  \bar \pi^{\rm T}_{i},~ik^i \bar \pi^{\rm T}_{i}=0.
\end{align}
However, it is important to note that the rotation group is doubled on the CTP, and the physical rotation group acting on $\bar\pi_i$ differs from that of a physical 3-vector.
As discussed in the previous sections, the physical rotation acting on $\bar g_{\mu\nu}$ is trivial, and thus $\bar\pi_i$ transforms as a scalar under spatial rotations.
In contrast, the physical rotation acting on $\pi^i$ is nonlinearly realized and corresponds to a constant shift.
Indeed, products involving $\pi^i$ and $\bar\pi_i$ appear in expressions such as $I_d^{(1,0,1)} \supset - \dot{\bar \pi}_i \dot\pi^i (\rho + P)$.
The shift symmetry in $\pi^i$ ensures that such building blocks are invariant under spatial rotations. 
Therefore, it is consistent to assign a \textit{linear} transformation property to $\pi^i$ and $\bar \pi_i$, allowing its spatial index to be treated as if it belongs to a linear representation of the rotation group.

With the scalar-vector decomposition, the effective action is rewritten to
\begin{align}
	I_{\pi}  = &\int \frac{d\omega}{2\pi} \int \frac{d^3k}{(2\pi)^3}
	\tilde{\bm{\pi}}^\dagger \tilde G^{-1} \tilde{\bm{\pi}}	,
\end{align}
where we introduced 
\begin{align}
	\tilde{\bm {\pi}} = \begin{pmatrix}
		\pi^{{\rm T}i}
		\\
		ik_i \pi^{\rm L}
		\\
		\bar \pi^{\rm T}_{i}
		\\
		ik_i \bar \pi^{\rm L}_{}
	\end{pmatrix},
\end{align}
and
\begin{align}
	\tilde G^{-1} = 
	\begin{pmatrix}
		0 & 0   & \tilde{\mathcal E}^{{\rm T}*} & 0 &
		\\
		0 & 0 & 0  & \tilde{\mathcal E}^{{\rm L}*} 
		\\
		\tilde{\mathcal E}^{{\rm T}} & 0  & \tilde {\mathcal N}^{\rm T}   & 0  
		\\
		0 & \tilde{\mathcal E}^{{\rm L}}  & 0 & \tilde {\mathcal N}^{\rm L } 
	\end{pmatrix}.
	\label{ntildematrix}
\end{align}
The transverse modes decouple from the longitudinal mode and $\theta$.
The dispersion relation for the retarded longitudinal and transverse modes are given by $\tilde {\mathcal E}^{\rm L}=0$ and $\tilde {\mathcal E}^{\rm T}=0$, respectively.

\subsection{Longitudinal mode for a neutral fluid}
\label{sec:nt}

Let us first consider the dispersion relation for the longitudinal mode for a neutral fluid.
This argument is included in the analysis for a general U(1) charged fluid in the next subsection.
However, the expressions are simpler than the charged case, and it may be useful to compare the present analysis with the previous results.
$\tilde{\mathcal E}^{\rm L}=0$ is recast into a well-known form 
\begin{align}
0&=
\kappa_{2} \epsilon \omega ^4 +i( \beta  \kappa_{2} \rho_\beta - \epsilon  (\rho+P))\omega^3
\notag 
\\
&
+ \left(k^2 \left(-\kappa_{1} \lambda_{1}-\lambda_{2} (\kappa_{2}+\lambda_{1})+\epsilon  \left(\zeta +\frac{4 \eta }{3}\right)\right)+\beta  \rho_\beta  (\rho+P)\right)\omega ^2
\notag 
\\
&+\frac{1}{3} i k^2  (\beta  \rho_\beta  (3 \zeta +4 \eta )+3 \lambda_{2} (\rho+P)-3 \beta  P_\beta  (\kappa_{2}+\lambda_{1}))\omega 
\notag 
\\
&
+\frac{1}{3} k^4 (3 \zeta +4 \eta ) \kappa_{1}-\beta  k^2 P_\beta  (\rho+P)
.\label{dispfull}
\end{align}
Eq.~\eqref{dispfull} has been discussed in various seminal works in the past.
As our approach significantly differs from others, checking the consistency of this result can be a benchmark of our detailed analysis.
In fact, the same equation is found in Eq.~(4.8) in a seminal work~\cite{Kovtun:2019hdm}.
Although it was derived in a different contexts, we exactly reproduced the same result by a careful matching of conventions and use of thermodynamic relations~\footnote{Readers can check the consistency with the help of the footnote 8 in Ref.~\cite{Kovtun:2019hdm}, and our translation table~\ref{tab1}. }.

Note that we have yet to impose the KMS condition.
For nonzero $\kappa_2$ and $\epsilon$, the dispersion relation is a fourth-order algebraic equation.
One often finds the solutions using the solution formula, which leads to unstable gapped excitation when the KMS condition is imposed:
\begin{align}
	{\rm Im}~ \omega \propto \kappa_2^{-1} \geq 0.\label{wrong}
\end{align}
While the mode is stable for the exact $\kappa_2 =0$, the instability gets worse as $\kappa_2\to 0$.
It was said that the $\kappa_2=0$ is a ``peculiar singular limit''~\cite{Hiscock:1985zz}. 
This is known as hydrodynamical instability.
However, such a pathological behavior is simply because of the inconsistent treatment of the gradient expansion, and Eq.~\eqref{wrong} is outside the allowed parameter space in the effective theory.
In fact, $\mathcal O(\varepsilon^{n>1})$ in equation of motion is understood as a higher-order derivative correction; therefore, we must treat them perturbatively~\cite{Abbasi:2017tea}.
To be more specific, when solving the constraint equation 
\begin{align}
	\bar \pi^*_0  =\bar \pi_i \frac{ \mathcal E^i{}_0 }{\mathcal E^0{}_0 },
\end{align}
to find Eq.~\eqref{dispfull}, we need to be more careful.
The dispersion relation is found by simplifying 
\begin{align}
	k_ik^j\left(\mathcal E^i{}_j + \frac{ \mathcal E^i{}_0 \mathcal E^0{}_j }{\mathcal E^0{}_0 } \right) =0.
\end{align}
At the level of EoM, one may multiply $\mathcal E^0{}_0$ to eliminate the denominator and find Eq.~\eqref{dispfull}.
However, we cannot justify the prescription at the action level.
Instead, we rewrite the denominator as
\begin{align}
	\frac{1}{2\mathcal E^0{}_0} = \frac{1}{ - \beta \rho_\beta \omega ^2 + i \omega ^3 \epsilon  
+ik^2 \omega  \kappa  } = \frac{1}{- \beta \rho_\beta \omega ^2 }\left( \frac{1}{1 - \frac{i \omega ^3 \epsilon  
+ik^2 \omega  \kappa}{\beta \rho_\beta \omega ^2 }}\right).
\end{align}
As we mentioned below Eq.~\eqref{deffirstsigma}, the gradient expansion is the expansion in $\varepsilon = (\rho+P)^{-1}v \nabla $ with the viscosity coefficients $v=\epsilon,\lambda,\cdots$.
We have to truncate the expansion at the first order in $\mathcal O(\varepsilon)$:
\begin{align}
	\frac{1}{1 - \frac{i \omega ^3 \epsilon  
+ik^2 \omega  \kappa}{\beta \rho_\beta \omega ^2 }} = 1 +  \frac{i \omega ^3 \epsilon  
+ik^2 \omega  \kappa}{\beta \rho_\beta \omega ^2 } +\mathcal O(\varepsilon^2).
\end{align}
Thus, while we truncated the derivative expansion at first order in the constitutive relation, the higher order derivatives enter the dispersion relation for the NG modes when solving the constraint equations.
In fact, in Landau frame where $\kappa =\epsilon =0$, we can safely solve the constraint equation without issue, and the derivative expansion scheme in the dispersion relation is identical to that of the constitutive relation.
However, the mode analysis is reliable only up to $\mathcal O(\varepsilon^1)$ in a general hydrodynamical frame.
We must further expand the dispersion relation to the same order of the constitutive relation.
Hence, we rather find
\begin{align}
0&=
i( \beta  \kappa_{2} \rho_\beta - \epsilon  (\rho+P))\omega^3
+ \left(\beta  \rho_\beta  (\rho+P)\right)\omega ^2
\notag 
\\
&+\frac{1}{3} i k^2  (\beta  \rho_\beta  (3 \zeta +4 \eta )+3 \lambda_{2} (\rho+P)-3 \beta  P_\beta  (\kappa_{2}+\lambda_{1}))\omega 
\notag 
\\
&
-\beta  k^2 P_\beta  (\rho+P) + \mathcal O(\varepsilon^2)
.\label{dispfull2}
\end{align}
The difference of Eq.~\eqref{dispfull} and \eqref{dispfull2} is unreliable correction in $\varepsilon$ as we only expand the constitutive relation to first order, and thus we must drop for a consistent analysis in effective theory point of view.

\medskip
Next let us consider the mode analysis of Eq.~\eqref{dispfull2}.
Naive mistake is to take the zero of this third order algebraic equation.
One finds three roots in this analysis.
One of the three modes is in fact unphysical, since it is outside the convergence radius of the derivative expansion.
The truncation of this solution is not ad-hoc, as reliability of Eq.~\eqref{dispfull2} is not valid.
The consistent mode analysis should be given by expanding the solution to the series in the expansion parameter $\varepsilon$:
\begin{align}
	\omega = \sum_{n=0}^\infty \varepsilon^n \omega_n.
\end{align} 
Below we determine $\omega_n$ order by order in $\varepsilon$.
The lowest-order solution is immediately found as
\begin{align}
	\omega_0 = \pm c_sk,  \label{1stdisp}
\end{align}
where 
\begin{align}
	c_s^2 = \frac{\partial P}{\partial \rho} = \frac{P_\beta}{\rho_\beta }. \label{cddef1}
\end{align}
$c_s^2$ is determined by zeroth-order hydrodynamics, i.e., thermodynamics, which is undefined at this order.
This mode result implies the obvious fact that we have the free propagating sounds if the first order coefficients are small.
The dispersion relation at first order is solved as
\begin{align}
	\omega_1 = -\frac{2 i k^2 \eta  }{3 (\rho+P)}-\frac{i k^2}{2 (\rho+P)}
	\left(\zeta - \frac{\partial P}{\partial \rho}\lambda_{1}\right)
	-\frac{i k^2  }{2 \beta  P_\beta   }\frac{\partial P}{\partial \rho} \left ( \lambda_{2}
	-\frac{\partial P}{\partial \rho} \epsilon\right) .\label{omega1neutral1}
\end{align}
Thus, $\kappa_i$ disappears from the first-order analysis.
This is reasonable as $\kappa_i$ is redundant freedom for a general neutral fluid: one can eliminate $\kappa_i$ by a field redefinition in the absence of a number flux.  
Note that we used Eq.~\eqref{1stdisp}, which corresponds to using the zeroth order hydrodynamics, just like using the on-shell solutions.
However, there is no ambiguity in the use of Eq.~\eqref{1stdisp}.
The hydrodynamical frame invariants $\tilde \zeta$ and $\tilde \lambda$ write $\omega_1$, so the mode analysis does not depend on the hydrodynamical frame choice.
Imposing the KMS condition, one can write Eq.~\eqref{omega1neutral1} as
\begin{align}
	\omega_1 = -\frac{2 i k^2 \eta  }{3 (\rho+P)}-\frac{i k^2}{2 (\rho+P)}
	\left(\zeta - 2 \frac{\partial P}{\partial \rho}\lambda
	+ \left(\frac{\partial P}{\partial \rho}\right)^2 \epsilon\right) .\label{omega1neutral}
\end{align}
Now the unitarity constraint~\eqref{positivdefeplam} implies that ${\rm Im}~ \omega_1 \leq 0$ if $\rho+P> 0$ is satisfied.
The sound speed is not affected by the first order coefficients as ${\rm Re}~ \omega_1 =0$.
We extend this result to a general U(1) charged fluid shortly.

We did not observe gapped excitations as in the case of other mode analyses in open systems~\cite{Minami:2015uzo, Hongo:2018ant}.
This is because the hierarchy in $\varepsilon$ dominates over the dispersion relation: mixing of $\mathcal O(\varepsilon^0)$ and $\mathcal O(\varepsilon^1)$ is not allowed.
Gapped mode might be found in the regime, $k \ll  \omega $.
If we naively solve the dispersion relation~\eqref{dispfull2} with 
\begin{align}
	\omega = \omega_0 + \vartheta \omega_1 + \mathcal O(\vartheta^2),~k = \vartheta k, \label{mdpt3}
\end{align}
one finds a hydrodynamical frame-dependent gapped mode, $\omega_0 = \mathcal O(\varepsilon^{-1})$.
Mathematically, this solution is found when the two terms in the first line in Eq.~\eqref{dispfull2} balance, which is the invalid regime in the gradient expansion.
The fact that a gap does not appear implies that the NG modes are massless and that dissipation is always higher order in the derivative expansion.
Thus, the effective action of the hydrodynamical NG modes is subject to extra constraints compared to that of a general QFT. 

$P\to 0$ limit corresponds to the nonrelativistic regime of the UV sector.
In this case, we find $\omega_0 = 0$, and then 
\begin{align}
	\omega_1 = - \frac{ik^2 (3\rho\lambda_2  +(3\zeta +4\eta) \rho_\beta) }{3 \rho \beta \rho_\beta},
\end{align}
which can be also found when taking $P\to 0$ and $P_\beta \to 0$ in Eq.~\eqref{omega1neutral1}.
Thus, for a vanishing pressure limit, the NG modes are purely diffusive.
$\omega_1$ depends on $\lambda_2$ and $\zeta$ in this limit, which are now hydrodynamical invariants as $\tilde \lambda \to \lambda_2$ and $\tilde \zeta \to \zeta$ in $P\to 0$ limit.
$\lambda_1$ and $\epsilon$ are eliminated by field redefinition, and the nontrivial first-order correction in the energy-momentum tensor is only spatial.

\subsection{Longitudinal mode for a general charged fluid}

For a general charged fluid, the mode analysis is more complicated than the neutral case.
The dispersion relation is now 6th order algebraic equation~(for a conformal U(1) charged fluid, see Ref.~\cite{Taghinavaz:2020axp})
\begin{align}
	0 =  \sum_{n=0}^6 d_n \omega^n, 
\end{align}
where we defined
\begin{align}
	d_0 & = \beta ^2 k^6 (3 \zeta +4 \eta ) (\kappa_{1} \chi_{1}-\tau_{1} \tau_{3})
	\notag 
	\\
	&+k^4(P_\mu  (3 \kappa_{1} n-3 \beta  \tau_{1} (\rho+P ))
	\notag 
	\\
	&
	\left.+(\beta  P_\beta -\mu  P_\mu ) \left(3 \beta  n \tau_{3}-3 \beta ^2 \chi_{1} (\rho+P )\right)\right),
\\
	d_1 & = 
	k^2((\beta  P_\beta -\mu  P_\mu ) (3 i n_\mu (\rho+P )-3 i n \rho_\mu )
	\notag 
	\\
	&
	+P_\mu  (3 i n (\beta  \rho_\beta -\mu  \rho_\mu )-3 i (\rho+P ) (\beta  n_\beta-\mu  n_\mu)))
	\notag 
	\\
	&
	+k^4(3 i \beta  n (\kappa_{1} \nu_{5}-\lambda_{2} \tau_{3})+3 i \beta  P_\mu  (\kappa_{1} (\nu_{2}+\tau_{2})-\tau_{1} (\kappa_{2}+\lambda_{1}))
	\notag 
	\\
	&
	+3 i \beta ^2 (\beta  P_\beta -\mu  P_\mu ) (\tau_{3} (\nu_{2}+\tau_{2})-\chi_{1} (\kappa_{2}+\lambda_{1}))
	\notag
	\\
	&
	-3 i \beta ^2 (\rho+P ) (\nu_{5} \tau_{1}-\lambda_{2} \chi_{1})
	\notag 
	\\
	&
	\left.-i (3 \zeta +4 \eta ) \left(\beta  \left(\beta ^2 (-\rho_\beta ) \chi_{1}+\beta  \mu  \rho_\mu  \chi_{1}+\beta  n_\beta \tau_{3}-\rho_\mu  \tau_{1}\right)+n_\mu (\kappa_{1}-\beta  \mu  \tau_{3})\right)\right)
\\
	d_2 & = 
	k^4\left(\beta ^2 (3 \zeta +4 \eta ) (\kappa_{1} \nu_{3}-\nu_{1} \tau_{3}-\nu_{4} \tau_{1}+\chi_{1} \epsilon )\right.
	\notag 
	\\
	&
	+3\beta ^2(\kappa_{2} \nu_{5} \tau_{1}+\lambda_{1} \nu_{5} \tau_{1}+\lambda_{1} \tau_{1} \tau_{3}+\lambda_{2} \nu_{2} \tau_{3}
	\notag
	\\
	&
	-\kappa_{1} (\lambda_{1} \chi_{1}+\nu_{5} (\nu_{2}+\tau_{2}))-\lambda_{2} \chi_{1} (\kappa_{2}+\lambda_{1})++\lambda_{2} \tau_{2} \tau_{3}))
	\notag 
	\\
	&
	+k^2(\beta  (3 \zeta +4 \eta ) (n_\mu \rho_\beta -n_\beta \rho_\mu )+P_\mu (3 n \epsilon +3 \beta  n_\beta (\kappa_{2}+\lambda_{1})
	\notag 
	\\
	&
	-3 (\beta  (\nu_{2}+\tau_{2}) (\beta  \rho_\beta -\mu  \rho_\mu )+\mu  n_\mu (\kappa_{2}+\lambda_{1})+\beta  \nu_{1} (\rho+P )))
	\notag 
	\\
	&
	+(\beta  P_\beta -\mu  P_\mu ) (3 \beta  \nu_{4} n-3 (n_\mu (\kappa_{2}+\lambda_{1})+\beta  (\beta  \nu_{3} (\rho+P )-\rho_\mu  (\nu_{2}+\tau_{2}))))
	\notag 
	\\
	&
	-3 n (\beta  (\nu_{5}+\tau_{3}) (\beta  \rho_\beta -\mu  \rho_\mu )+\rho_\mu  (\kappa_{1}+\lambda_{2}))
	\notag 
	\\
	&
	+3 (\rho+P ) (\beta  (\beta  \chi_{1} (\beta  \rho_\beta -\mu  \rho_\mu )+\beta  \nu_{5} n_\beta+\rho_\mu  \tau_{1})+n_\mu (\lambda_{2}-\beta  \mu  \nu_{5})))
\\
	d_3 & =
	3 i (\rho+P ) (\beta  n_\beta \rho_\mu -\beta  n_\mu \rho_\beta )
	\notag 
	\\
	&
	+k^2(-i (3 \zeta +4 \eta ) (\beta  (\beta  \nu_{3} (\mu  \rho_\mu -\beta  \rho_\beta )-\nu_{1} \rho_\mu +\beta  \nu_{4} n_\beta)+n_\mu (\epsilon -\beta  \mu  \nu_{4}))
	\notag 
	\\
	&
	+3 i \beta  P_\mu  (\epsilon  (\nu_{2}+\tau_{2})-\nu_{1} (\kappa_{2}+\lambda_{1}))
	\notag 
	\\
	&
	-3 i \beta ^2 (\beta  P_\beta -\mu  P_\mu ) (\nu_{3} (\kappa_{2}+\lambda_{1})-\nu_{4} (\nu_{2}+\tau_{2}))
	\notag 
	\\
	&
	+3 i \beta  n (\epsilon  (\nu_{5}+\tau_{3})-\nu_{4} (\kappa_{1}+\lambda_{2}))
	\notag 
	\\
	&
	+3i(n_\mu ((\kappa_{2}+\lambda_{1}) (\lambda_{2}-\beta  \mu  \nu_{5})-\beta  \lambda_{1} \mu  \tau_{3}+\kappa_{1} \lambda_{1})
	\notag
	\\
	&
	+\beta \left(\beta ^2 (-\nu_{2}) \nu_{5} \rho_\beta +\beta  \lambda_{2} \nu_{3} \rho -\beta  \nu_{1} \nu_{5} \rho -\kappa_{1} \nu_{2} \rho_\mu -\lambda_{2} \nu_{2} \rho_\mu \right.
	\notag 
	\\
	&
	\beta ^2 (-\nu_{5}) \rho_\beta  \tau_{2}++\beta  \mu  \nu_{2} \nu_{5} \rho_\mu +\beta  \nu_{4} \rho  \tau_{1}-\kappa_{1} \rho_\mu  \tau_{2}+\kappa_{2} \rho_\mu  \tau_{1}
	\notag 
	\\
	&
	\beta ^2 (-\nu_{2}) \rho_\beta  \tau_{3}+\beta  \mu  \nu_{2} \rho_\mu  \tau_{3}+\beta  \mu  \nu_{5} \rho_\mu  \tau_{2}-\lambda_{2} \rho_\mu  \tau_{2}
	\notag 
	\\
	&
	-\beta ^2 \rho_\beta  \tau_{2} \tau_{3}+\beta  \mu  \rho_\mu  \tau_{2} \tau_{3}+\beta  n_\beta (\nu_{5} (\kappa_{2}+\lambda_{1})+\lambda_{1} \tau_{3})
	\notag 
	\\
	&
	+\beta  \chi_{1} (\beta  \kappa_{2} \rho_\beta -\kappa_{2} \mu  \rho_\mu +\rho  (-\epsilon ))+\beta  \mathcal{P} (\lambda_{2} \nu_{3}-\nu_{1} \nu_{5}+\nu_{4} \tau_{1}-\chi_{1} \epsilon ))))
\\
	d_4 & =
	3 \beta  \kappa_{2} (n_\mu \rho_\beta -n_\beta \rho_\mu )+3\rho+P (n_\mu (\beta  \mu  \nu_{4}-\epsilon )
	\notag
	\\
	&
	+\beta  (\beta  \nu_{3} (\beta  \rho_\beta -\mu  \rho_\mu )+\nu_{1} \rho_\mu +\beta  \nu_{4} (-n_\beta)))
	\notag 
	\\
	&
	+k^2\left(\beta ^2 (3 \zeta +4 \eta ) (\nu_{3} \epsilon -\nu_{1} \nu_{4})\right.
	\notag 
	\\
	&
	+3\beta ^2(-\kappa_{1} \lambda_{1} \nu_{3}-\lambda_{1} \lambda_{2} \nu_{3}+\lambda_{1} \nu_{1} \nu_{5}+\lambda_{2} \nu_{2} \nu_{4}+\lambda_{2} \nu_{4} \tau_{2}-\nu_{2} \nu_{5} \epsilon -\nu_{5} \tau_{2} \epsilon
	\notag 
	\\
	&
	+\kappa_{1} \nu_{4} (\nu_{2}+\tau_{2})+\lambda_{1} \nu_{1} \tau_{3}+\kappa_{2} (-\lambda_{2} \nu_{3}+\nu_{1} \nu_{5}-\nu_{4} \tau_{1}+\chi_{1} \epsilon )-\nu_{2} \tau_{3} \epsilon -\tau_{2} \tau_{3} \epsilon ))
\\
	d_5 & =
	-3 i \beta ^2 (\rho+P ) (\nu_{3} \epsilon -\nu_{1} \nu_{4})-3i\kappa_{2}(n_\mu (\epsilon -\beta  \mu  \nu_{4})
	\notag 
	\\
	&
	+\beta  (\beta  \nu_{3} (\mu  \rho_\mu -\beta  \rho_\beta )-\nu_{1} \rho_\mu +\beta  \nu_{4} n_\beta))
\\
	d_6 & = 3 \beta ^2 \kappa_2 (\nu_3 \epsilon -\nu_1 \nu_4)
\end{align}
As in the case of the neutral fluid, the above dispersion relation is reliable up to $\mathcal O(\varepsilon)$, and we need to solve the dispersion relation order by order in $\varepsilon$.
In the low energy limit, the zeroth order solution is found as
\begin{align}
	\omega_0 = \pm c_{s/n} k,~c_{s/n}^2 \equiv \left( \frac{\partial P}{\partial \rho} \right)_{s/n}.\label{adibacs}
\end{align}
The adiabatic sound velocity $c_{s/n}$ contains Eq.~\eqref{cddef1} in the neutral limit.
The explicit form of $c_{s/n}$ is found in Eq.~\eqref{exp:adiabaticsv}.
Zeroth order thermodynamics is stable if $c_{s/n}^2\geq 0$.
The first-order solution is found as
\begin{align}
	\omega_1 &= - i \Gamma_{\rm L} k^2  ,\label{adibacs2}
	\\
	 \Gamma_{\rm L} &\equiv \frac{1 }{2 (\rho +P)}\left[	\frac{4}{3}\eta+ {\bm \zeta} + f^2  \left( \frac{\partial P}{\partial \rho} \right)_{s/n} {\bm \chi}_1 + \beta \left(\frac{\partial P}{\partial n}\right)_\rho {\bm \tau} \right],\label{gameq}
\end{align}
where ${\bm \zeta}$, ${\bm \chi}_1$ and ${\bm \tau}$ are defined in Sec.~\ref{fielddef}, and we defined
\begin{align}
	f^2 &= \frac{\beta ^2 (\rho+P )^2 ((\rho+P-\mu n) (\beta  n_\beta-\mu  n_\mu)-\beta  n \rho_\beta )^2}{\left(n_\mu (\rho+P-\mu n)^2-\beta  n (n_\beta (\mu  n-2 (\rho+P) )+n \rho_\beta )\right)^2}.
\end{align}
This is the main result of this paper.
Although Eq.~\eqref{gameq} is written in a simple form, concrete expressions for $ \left( \frac{\partial P}{\partial \rho} \right)_{s/n}$ and $\left(\frac{\partial P}{\partial n}\right)_\rho $ are tedious, as shown in App.~\ref{appadsv}.
Note that $f^2\geq 0$ since the thermodynamic variables are real.
${\bm \chi}_2$ drops from the first order solution.
As the bold faces are hydrodynamical frame invariants, the mode analysis does not depend on the hydrodynamical frame choice. 
We have substituted $\omega_0$ into the first-order dispersion relation to find $\omega_1$, which corresponds to solving the on-shell hydrodynamics \textit{for NG modes}.
This is why the boldfaces write the first-order dispersion relation.
Importantly, there is no ambiguity in the use of the on-shell condition, in contrast to the case in Sec.~\ref{fielddef}.
As discussed in Sec.~\ref{fielddef}, the KMS condition and unitarity of the UV system implies ${\bm \chi} \geq 0$, ${\bm \tau} = 0$ and ${\bm \zeta}\geq 0$.
Therefore, $\Gamma_{\rm L}\geq 0$ and stable if $\rho+P>0$.
Note that we have not taken any specific hydrodynamical frame for writing down the effective action for the NG modes, and hence, this mode analysis applies to the Eckart frame as well.

For $\rho+P =0$, we find
\begin{align}
	\tilde{\mathcal E}^{\rm L} =k^3 \left(  \frac{4}{3} \eta +  \zeta +\lambda_1 +\frac{n (\lambda_2+\epsilon )}{\beta  n_\beta-\mu  n_\mu}\right) + \mathcal O(\varepsilon^2).
\end{align}
As $\tilde \zeta = \zeta +\lambda_1$ and $\tilde \lambda = \lambda_2+\epsilon$ when $\rho+P=0$, the result is field redefinition invariant.
Thus, $\omega$ disappears from the dispersion relation so that the dynamical sound waves do not exist.
Then, $\rho+P \geq  0$ guarantees the stability of first-order hydrodynamics in thermal states.
The adiabatic sound velocity is solely determined by its thermodynamic nature, which, like the neutral case, does not depend on first-order hydrodynamics.

\subsection{Transverse mode}

The transverse mode is absent in a perfect fluid, as sound waves are ``longitudinal'' as we study in high school.
However, in a viscous fluid, the situation is slightly different.
Just like the longitudinal mode, the dispersion relation for the transverse mode is~\cite{Hiscock:1985zz} 
\begin{align}
	\omega (\rho +P) + i\kappa_2 \omega^2+i \eta  k^2 =0.
\end{align}
We immediately find $\omega_0=0$, and then the first order correction is
\begin{align}
	\omega_1 = - i \Gamma_{\rm T} k^2 ,~\Gamma_{\rm T} \equiv  \frac{\eta}{\rho +P}.\label{transmdan1}
\end{align}
Thus, $\kappa_i$ disappears in the first order, and the dispersion relation is hydrodynamical frame invariant.
Since $\eta\geq 0$, ${\rm Im}~ \omega <0$ if $\rho+P>0$. 
The transverse mode decouples from $\theta$, so the conclusion does not depend on whether charged or not.
For $\rho+P=0$, we find $\kappa$ hydrodynamical frame independent.
$\kappa \neq 0$ is a peculiar situation, and the transverse mode is unstable unless $\eta = 0$.
However, such a solution needs to be consistent with the perturbative expansion of $\omega$.
Hence, the transverse mode does not exist for $\rho+P=0$.
Thus, the $\rho+P\geq 0$ is also a necessary condition for stability in the transverse mode.

\subsection{Mode analysis in non-comoving inertial frames}

So far we assumed a specific inertial frame where the fluid is comoving.
The original UV theory enjoys the full Poincaré symmetry in flat spacetime, including Lorentz boosts.
The presence of $\beta^\mu$ in the ensemble spontaneously breaks this symmetry.
Then, a natural question is if the mode analysis in a non-comoving inertial frame is equivalent to that in the rest frame.

The answer to this question is straightforward as we employed the nonlinear realization theory.
The broken boost symmetry is nonlinearly realized by the NG field $\pi^\mu$ through Eq.~\eqref{def:nontl}. 
Therefore, the dynamical theory of hydrodynamical modes are manifestly covariant under the nonlinear Lorentz boost.

To illustrate this more clearly, we consider an infinitesimal Lorentz boost in the $\tau$–$x$ plane, generated by a tangent vector $\xi^\mu = v(x\delta^\mu_0 + \tau\delta^\mu_1)$. 
Under this transformation, the Minkowski background remains invariant, while the NG fields transform as $\pi^0 \to \tilde{\pi}^0 = \pi^0 + v x$ and $\pi^1 \to \tilde{\pi}^1 = \pi^1 + v \tau$. 
Although $\dot{\tilde{\pi}}^0 = \dot{\pi}^0$, the time derivative of the spatial component changes: $\dot{\tilde{\pi}}^1 \neq \dot{\pi}^1$. 
Consequently, the effective action in the boosted frame is invariant when expressed in terms of $\tilde{\pi}^\mu$.

This nonlinear realization implies that the dynamics of sound waves in a fluid at rest are physically equivalent to those in a boosted frame.
More precisely, after the boost transformation, one identifies the fluid comoving with the new inertial frame, and hydrodynamical modes defined in that fluid is physically equivalent to the original modes.
In this sense, the theory of hydrodynamical modes is always defined in a comoving inertial frame.
This contrasts with a naive application of Lorentz boosts to the wavevector $k_\mu$.
While an identification by the boost is intuitive, it is only physically meaningful when the field variables are linear representations of Lorentz boosts.

The theory formulated using $\pi^\mu$ directly in the boosted frame was analyzed in Ref.~\cite{Hiscock:1985zz} in a traditional perturbation theory, and then instabilities were reported. 
It is now clear that this formulation is not dynamically equivalent to that in the rest frame, and the emergence of instabilities is not surprising.
The dynamically equivalent theory in the boosted frame is instead formulated in terms of $\tilde{\pi}^\mu$, whose mode structure is manifestly identical to that in the original rest frame.
The theory of $\pi^\mu$ in the non-comoving frame mixes density waves with mass transfer due to the moving frame, leading to a misidentification of the dynamical degrees of freedom.

%The origin of unstable modes in non-comoving inertial frames can be further clarified.
%Since hydrodynamics is defined through symmetry principles in the fluid spacetime~\cite{Crossley:2015evo}, any perturbative analysis in a non-comoving inertial frame must refer back to this structure. 
%In a boosted configuration, the thermal vector takes the form
%\begin{align}
%	\beta^\mu = \beta_c(\delta^\mu_0 + v^i \delta^\mu_i + \dot \pi^\mu).
%\end{align}
%Consider a diffeomorphism transformation from the fluid spacetime to this frame by $\xi$, which is found as a solution of
%\begin{align}
%	v^i \delta^\mu_i + \dot \pi^\mu = \dot \xi^\mu 
%\end{align}

%The coordinate transformation from the unitary gauge is then formally written as
%\begin{align}
%	\xi^\mu = - \frac{\partial_0}{\partial_0 + v^i \partial_i} \pi^\mu,
%\end{align}
%via Eq.~\eqref{embed}.
%This expression highlights that the gauge transformation becomes nonlocal in a boosted frame, reflecting the conflation of mass transfer by fast modes and hydrodynamical modes as density waves.
%Consequently, the embedding in Eq.~\eqref{def:nontl} acquires nonlocality, and the resulting effective theory no longer possesses a well-defined local dynamical structure. 
%Therefore, the observed instabilities and acausal behavior can be attributed not to fundamental pathologies of hydrodynamics, but rather to a misidentification of the physical degrees of freedom in the boosted frame. 

\paragraph{Section summary.}
This section presents the main contribution of this paper.  
Building on the effective action derived in the previous section, we performed a mode analysis of the hydrodynamical excitations.  
We carefully eliminated the redundancy in the NG modes by solving the constraint equations up to noise order, where noise fields correspond to the advanced-sector counterparts in the Schwinger-Keldysh formalism.  
The analysis was carried out within a consistent derivative expansion scheme.  
We emphasized that the derivative expansion of the constitutive relations is not, in general, aligned with that of the NG mode dynamics under generic hydrodynamical frames.  
A consistent truncation of derivative terms yields a manifestly frame-invariant mode structure, which, for linear modes, coincides with the Landau frame result.  
Upon imposing the KMS and unitarity conditions, we demonstrated that all hydrodynamical modes remain stable under the physical condition $\rho + P > 0$.

\section{Green functions and causality}
\label{gfsec}

In the previous section, we only focused on classical dynamics, the first-order action with respect to the noise field.
In this section, we extend the analysis to the semiclassical order, and we discuss the properties of Green functions.

This is the benefit of our formulation.
We have derived the noise action in a general hydrodynamic frame, and carefully eliminate the redundant variables by solving the constraint equations.
Consequently, we have the noise action written by the hydrodynamical frame invariant.
To be more specific, the linear analysis can determine the advanced and retarded propagators, $G_A$ and $G_R$.
In previous works, one may find the Keldysh propagator $G_K$ by the KMS condition.
In our framework, we can find $G_K$ from the second order action, and the KMS condition among the propagators are derived as discussed shortly.

\subsection{Green functions}

One finds the reduced noise matrix~\eqref{deftildeN} by eliminating the redundant freedom.
The matrix components in Eq.~\eqref{ntildematrix} are found as
\begin{align}
	\tilde {\mathcal N}^{\rm L} = \frac{ik^2(\rho+P)\Gamma_{\rm L} }{2\beta}+ \mathcal O(\varepsilon^2),
\end{align}
where we use $\omega^2 = c_{s/n}^2 k^2 + \mathcal O(\varepsilon^1)$, and 
\begin{align}
	\tilde {\mathcal N}^{\rm T} &= \frac{ik^2(\rho + P)\Gamma_{\rm T} }{4\beta}+ \mathcal O(\varepsilon^2),
\end{align}
with $\omega = 0+ \mathcal O(\varepsilon^1)$.
The equations of motion in the gradient expansion are
\begin{align}
		\tilde {\mathcal E}^{\rm L} &=\frac{1}{2} (\rho+P) (\omega - c_{s/n} k + i \Gamma_{\rm L} k^2 )(\omega + c_{s/n} k + i \Gamma_{\rm L} k^2 ) + \mathcal O(\varepsilon^2),
		\\
		\tilde {\mathcal E}^{\rm T} &=\frac{1}{2} (\rho+P) (\omega  + i \Gamma k^2 ) + \mathcal O(\varepsilon^2).
\end{align}
Note that the substitution of the leading order dispersion relation implies the on-shell condition, including noise.
Now we find the reduced FDR:
\begin{align}
	\frac{\tilde {\mathcal N}}{\tilde{\mathcal E}\tilde{\mathcal E}^*}= \frac{2}{\beta \omega} \left(\frac{1}{\tilde{\mathcal E}^*} - \frac{1}{\tilde{\mathcal E}}\right).\label{EK}
\end{align}
Thus, the KMS condition for the full action guarantees the KMS condition for the dynamical fields after solving the constraint equations.
%Eq.~\eqref{EK} is understood as another representation of the dynamical KMS condition proposed in Ref.~\cite{Glorioso:2017fpd}.
$\tilde G^{-1}$ is invertible and we find
\begin{align}
	G
	=
	\begin{pmatrix}
		-\tilde{\mathcal E}^{{\rm T}-1} \tilde {\mathcal N}^{\rm L} \tilde{\mathcal E}^{{\rm L}*-1} & 0 &  \tilde{\mathcal E}^{{\rm L}-1} & 0
		\\
		0 & -\tilde{\mathcal E}^{{\rm L}-1} \tilde {\mathcal N}^{\rm L} \tilde{\mathcal E}^{{\rm L}*-1} & 0 & \tilde{\mathcal E}^{{\rm T}-1}
		\\
		\tilde{\mathcal E}^{{\rm T}*-1} & 0 & 0 & 0
		\\
		0 & \tilde{\mathcal E}^{{\rm L}*-1} & 0 & 0
	\end{pmatrix}
	\equiv 
	\begin{pmatrix}
		G_K & G_R
		\\
		G_A & 0
	\end{pmatrix}.
	\label{356}
\end{align}
$G_K$, $G_R$ and $G_A$ are the Keldysh~(statistical), retarded and advanced propagators, respectively.
The effective action for the NG mode coincides with the MSR action for the Brownian motion~\eqref{msrbp}.
By taking
\begin{align}
	m \to \frac{\rho+P}{2},~\omega_0^2 =c_{s/n}^2 k^2,~\gamma\to 2k^2 \Gamma_{\rm L},
\end{align}
in Eqs.~\eqref{gflgb} and \eqref{gktt'}, we find
\begin{align}
	G_R(t,k) &= \Theta(t)\frac{e^{-k^2 \Gamma_{\rm L}t }}{ c_{s/n}k}\sin\left( c_{s/n}k t \right)\label{gflgb1},
\\
	G_{K}(t,k) &=  \frac{2e^{-k^2 \Gamma_{\rm L} t}}{\beta (\rho +P)c_{s/n}^2 k^2} \left(\cos \left( c_{s/n} k t\right) + \frac{\Gamma_{\rm L} k}{ c_{s/n} } \sin \left( c_{s/n} k t \right)\right).\label{gktt'2}
\end{align}
Note that the dispersion relations are reliable only up to $\mathcal O(\varepsilon)$.
The retarded propagator is constrained to $t>0$.
Combining these Eqs.~\eqref{gflgb1} and \eqref{gktt'2} with Eqs.~\eqref{sous1} and \eqref{eq:338}, one can compute any correlation functions of $T$ and $J$.

These Green functions are identical to those found in Brownian motion bounded in a harmonic potential, with similar expressions derived in the context of hydrodynamics~\cite{Kovtun:2012rj}. 
However, this work explicitly confirmed the hydrodynamical frame invariance of these Green functions.

\subsection{Comments on causal structure}

Superluminal modes may arise when higher-order derivatives are introduced inconsistently. However, even when the effective theory is properly constructed, there remain two key aspects of causality to address.

We first consider the constraint on the adiabatic sound speed. Thermodynamic variables determine $c_{s/n}$, which is, in principle, unconstrained. While fast modes in theories with Lorentz boost symmetry cannot propagate faster than the speed of light, the superluminal propagation of sound waves is not necessarily forbidden, as it does not involve the transfer of fast modes themselves.

The stability condition $\Gamma_{\rm L} \geq 0$ ensures that $G_R(t) = 0$ for $t < 0$ when performing the inverse Fourier transform with respect to $\omega$. Mathematically, this condition is satisfied when the zeros of $\tilde{\mathcal{E}}^{\rm L}$ do not extend into the region ${\rm Im}~ \omega > 0$. For real $k$, we find
\begin{align}
	{\rm Im}~ \omega = -\Gamma_{\rm L}k^2 < 0. \label{const1}
\end{align}
The authors of Ref.~\cite{Heller:2022ejw} extended this inequality to complex wavenumbers, showing that ${\rm Im}~ \omega(k) \leq |{\rm Im}~ k|$ by analyzing the analytic structure of a stable retarded Green’s function. Generalizing Eq.~\eqref{const1}, they derived
\begin{align}
	{\rm Im}~ \left[\pm c_{s/n} k - i\Gamma_{\rm L}k^2 \right] < |{\rm Im}~ k|. \label{const22}
\end{align}
For $k = i p$ with $p \in \mathbb{R}$ and $p \to 0$, this yields $c_{s/n} \leq 1$. Since $\Gamma_{\rm L} \geq 0$ is guaranteed by unitarity, the KMS condition, and the null energy condition, the upper bound on the adiabatic sound speed in first-order hydrodynamics is supported by these principles. The connection between causality and stability is further discussed in Refs.~\cite{Olson:1990rzl,Bemfica:2020zjp,Heller:2022ejw,Gavassino:2023myj}.

\medskip
Next, we consider the causal structure with respect to the sound speed. For nontrivial dispersion relations, there is a subtlety in defining the propagation speed of waves and signals, such as the phase velocity, group velocity, front velocity, etc. In particular, the standard definition of front velocity is incompatible with hydrodynamics, as it breaks down in the $k \to \infty$ limit.

For $\Gamma_{\rm L} = 0$, the retarded Green’s function \eqref{gflgb1} reduces to that of a relativistic scalar field in Minkowski spacetime with $c = c_{s/n}$. The retarded propagator must vanish for spacelike separations with respect to $c_{s/n}$. In other words, defining $r_{s/n} \equiv c_{s/n} t - x$, we find that $G_R$ vanishes for $r_{s/n} < 0$, which defines the sound cone. This confirms that sound waves cannot propagate faster than the sound speed, ensuring causality with respect to the sound cone.

\medskip

For $\Gamma_{\rm L} > 0$, the sound wave becomes dissipative, and the notion of causality becomes more subtle. One can further Fourier transform the Green's functions in $k$:
\begin{align}
	G_R(x-y) &= \int \frac{d^3 k}{(2\pi)^3} e^{i \mathbf{k} \cdot (\mathbf{x} - \mathbf{y})} G_{R}(t,k) 
	\notag 
	\\
	&=\Theta(t) \int_0^{\infty} \frac{k^2 dk}{2\pi^2}\int_{-1}^1 \frac{d\lambda}{2} e^{ik\lambda |\mathbf{x}-\mathbf{y}|} \frac{e^{-k^2 \Gamma_{\rm L}t }}{ c_{s/n} k}\sin\left( c_{s/n} k t \right)
	\notag 
	\\
	&=\frac{\Theta(t)}{(4\pi)^{3/2} c_{s/n} |\mathbf{x}-\mathbf{y}| \sqrt{\Gamma_{\rm L} t}}  e^{-\frac{(c_{s/n} t-|\mathbf{x}-\mathbf{y}|)^2}{4\Gamma_{\rm L}t}}.     \label{Grrealsp}
\end{align}
In the $\Gamma_{\rm L} \to 0$ limit, the Gaussian distribution approaches a delta function, and the sound waves are constrained to the sound cone. However, a nonvanishing $\Gamma_{\rm L}$ implies that the retarded propagator becomes nonzero for $r_{s/n} < 0$, meaning the correlation function extends outside the sound cone. This is the well-known causality issue in first-order hydrodynamics. The Gaussian distribution has nonzero tails for $|\mathbf{x}-\mathbf{y}| \to \infty$, which may naively extend beyond the light cone. This behavior is evident even for transverse modes with zero sound speed, indicating that bounding $c_{s/n}$ is insufficient to resolve the issue.

However, it is evident that Eq.~\eqref{Grrealsp} has a limited validity range, as hydrodynamics is an effective theory based on the derivative expansion. In Eq.~\eqref{Grrealsp}, the integration is formally performed from $k = 0$ to $\infty$ to obtain the real-space expression, but the exponential damping is only valid at leading order and does not hold in the $k \to \infty$ limit. Thus, the superluminal Gaussian tail does not imply a physical violation of causality but rather reflects the breakdown of the effective theory at short distances.

Therefore, the support of the retarded propagator in real space cannot be interpreted literally as a measure of causality. The width of the wave packet grows as $\sqrt{\Gamma_{\rm L} t}$, corresponding to the MFP of the fast modes. Diffusion occurs due to the propagation of these fast modes, not due to sound waves, as they carry net mass. In fact, the Gaussian wave packet in Eq.~\eqref{Grrealsp} suggests that the typical travel distance of fast modes is set by the MFP in the rest frame of the sound wave.

The maximum speed of the fast modes is determined by the UV theory and must not exceed the speed of light. Thus, Eq.~\eqref{Grrealsp} cannot be naively extrapolated to $|\mathbf{x}-\mathbf{y}| \to \infty$. The center of the sound packet still travels along the sound cone. In the rest frame of the sound wave, the fast modes diffuse with a maximum speed $c_{\rm fast} \leq 1$. Therefore, the maximum front speed $c_{\rm front}$ of the diffused sound wave is given by the relativistic addition of $c_{s/n}$ and $c_{\rm fast}$:
\begin{align}
	c_{\rm front} = \frac{c_{\rm fast} + c_{s/n}}{1 + c_{\rm fast} c_{s/n}} \leq 1,
\end{align}
as dictated by special relativity. The front speed never exceeds the speed of light, thereby preserving causality with respect to the light cone, even though causality with respect to the sound cone may be violated.

\paragraph{Section summary.}
This section focused on the analysis at noise order in the effective action.  
We derived the Keldysh Green function from the effective action constructed in the previous sections.  
By consistently eliminating the redundancy of the NG modes up to quadratic order in the noise fields, the resulting noise action remains frame invariant.  
Upon imposing the KMS condition at the level of the effective action, we verified that the Green functions satisfy the standard KMS relation.  
We also examined the causal structure of the theory with respect to the sound cone.  
In particular, we discussed that the adiabatic sound speed is bounded by the speed of light.  
Furthermore, the front velocity of dissipative processes—defined as the relativistic composition of the rest-frame velocity and the mass transport velocity in that frame—is also constrained by the causal structure, ensuring that all signal propagation remains subluminal in hydrodynamics.

\section{Conclusions}

We have formulated a symmetry-based effective theory for relativistic hydrodynamics with a global $U(1)$ symmetry, valid up to semi-classical noise order. Our construction relies solely on the input of a local four-vector $\beta^\mu$, without assuming any thermodynamic relations at the outset. By imposing local KMS and unitarity conditions, we have systematically derived thermodynamic relations and fluctuation–dissipation constraints directly from the effective action.

Although the foundational structure of this construction was already presented in the seminal work~\cite{Crossley:2015evo}, a key contribution of this paper is the identification of the quadratic action for hydrodynamical modes in the most general $U(1)$ hydrodynamics. The hydrodynamical mode arises as a nonlinear realization of spacetime symmetries, which makes the covariance of perturbation theory manifest. Our treatment also consistently incorporates noise contributions.

We performed a mode analysis by carefully eliminating unphysical redundancies. A consistent truncation of derivative terms yields a manifestly hydrodynamic-frame-invariant mode structure, which reproduces the Landau frame result in the linear regime.  
Upon imposing the KMS and unitarity conditions, we demonstrated that all hydrodynamical modes remain stable under the physical condition $\rho + P > 0$. The dominant energy condition, or the positivity of the enthalpy density is the minimal requirement for the stability of first-order hydrodynamics in flat spacetime.

Some of the results obtained in this work were known from earlier analyses of hydrodynamical perturbations using traditional methods. However, our approach differs in several important aspects. For instance, although the stability of long-wavelength hydrodynamical modes is widely recognized, to the best of our knowledge, this is the first work to explicitly establish hydrodynamic-frame invariance for a generic $U(1)$ fluid using a consistent gradient expansion. We also found that higher-order gradients can enter the equations of motion through the solution of constraint equations, even if the constitutive relations are first order.

A key advantage of our framework is the use of techniques from nonlinear realization theory. The hydrodynamical action is constructed to be invariant under nonlinearly realized Lorentz boosts, ensuring that the stability analysis is valid in any Lorentz frame. In our formulation, Lorentz boosts are realized nonlinearly, which ensures that the adiabatic sound speed remains unchanged. This means that the intrinsic sound speed is always defined in the rest frame of the fluid. The apparent sound speed arising from fluid motion should be distinguished from the adiabatic sound speed. One might expect that sound waves before and after a Lorentz transformation are related through a transformation of the wavevector, $k_\mu \to k'_\mu =\Lambda_\mu{}^\nu k_\nu$. This expectation holds for linearly realized representations such as photons. In contrast, sound waves are nonlinear realizations, and identifying them through linear transformations mixes variables that are not physically equivalent from a covariant perspective. This is the reason why mode analysis in the traditional approach depends on the choice of Lorentz frames.

\medskip
A natural direction for future work is to study the nonlinear evolution of Nambu–Goldstone modes. Since the first-order theory is now well-defined within the hydrodynamical regime, an extension to loop-level analysis becomes straightforward. Because the Green's functions are hydrodynamic-frame invariant, loop corrections will also be invariant if the interactions are constructed to respect the same invariance. This aspect will be discussed in future work.

In this paper, we applied the principle of hydrodynamic-frame independence to guide our mode analysis. Within the hydrodynamical regime, this approach proved successful. As a result, our perspective on first-order hydrodynamics differs from that of the BDNK theory~\cite{Bemfica:2017wps, Kovtun:2019hdm, Bemfica:2019knx, Bemfica:2020zjp}, where frame dependence plays a central role. A more detailed comparison between these approaches will be important in future studies.

Finally, we comment on possible applications of our effective field theory of dissipative fluids. Our effective action shares structural similarities with that of the effective field theory of inflation~\cite{Cheung:2007st}, and with its extensions to open systems~\cite{Hongo:2018ant, Salcedo:2024smn}. The effective theory of sound modes presented here may also be useful for describing warm inflation~\cite{Berera:1995wh,Berera:1995ie,Berera:2023liv}. 
It is straightforward to generalize our framework to cosmological backgrounds, such as Schwarzschild or Friedmann–Lemaître–Robertson–Walker spacetimes, and to couple it to the dynamical metric degrees of freedom. 
Our approach may be particularly suited to an action-based formulation of the effective field theory of large-scale structure~\cite{Carrasco:2012cv,Porto:2013qua}, since it does not rely on thermal equilibrium assumptions at the outset.

\acknowledgments

The author would like to thank Tsutomu Kobayashi, Navid Abbasi, Yoshimasa Hidaka, Masaru Hongo, Toshifumi Noumi, and Masahide Yamaguchi for their helpful discussions. This work was supported in part by the National Natural Science Foundation of China under Grant No. 12347101.

\appendix

\section{Notations}
\label{notationdiff}

As relativistic hydrodynamics is interdisciplinary and uses many symbols to describe in a general setup, depending on the contexts and authors, equivalent expressions are expressed in multiple ways.
This appendix provides a table of translation of hydrodynamical coefficients between our work and a seminal reference~\cite{Kovtun:2019hdm}.
We summarize the correspondence of variables and transport coefficients in Table~\ref{tab1}. This allows for a direct comparison of constitutive relations and facilitates the identification of frame-invariant quantities.
Here, the frame invariants such as $f_i$ and $\ell_i$ are defined in Ref.~\cite{Kovtun:2019hdm} as combinations of transport coefficients that are invariant under first-order field redefinitions of hydrodynamical variables. See Eq.~(2.10) of Ref.~\cite{Kovtun:2019hdm}.
Note that the external vector field is not introduced in Ref.~\cite{Kovtun:2019hdm}.

\begin{table}[htbp]
  \centering
  \caption{Correspondence of first-order hydrodynamical coefficients. Frame invariants (on-shell) means combinations relevant for observable response functions after eliminating redundant degrees of freedom using the equations of motion.}\label{tab1}
  \begin{tabular}{l c c}
    \hline
    \textbf{Physical quantity} & \textbf{Our notation} & Ref.~\cite{Kovtun:2019hdm}  \\
    \hline
    Temperature & $\beta^{-1}$ & $T$ \\
    Chemical potential & $\mu$ & $\mu$ \\
    4-velocity & $u^\mu$ & $u^\mu$ \\
    Adiabatic sound speed & $c_s,c_{s/n}$ & $v_s$\\
    \hline
    Spacetime metric & $g_{\mu\nu}$ & $\eta_{\mu\nu}$~(Minkowski) \\
    Induced metric & $\gamma_{\mu\nu}$ & $\Delta_{\mu\nu}$ \\ 
    Vector field & $A_\mu$ & {\rm none} \\ 
    \hline
    Energy density &  $(\rho, \epsilon,\lambda_1,\nu_4)$ & $(\epsilon, -\varepsilon_1,-\varepsilon_2,-\varepsilon_3)$ \\
    Pressure &  $(P,\lambda_2,\zeta,\nu_5)$ & $(p,-\pi_1,-\pi_2,-\pi_3)$ \\
    Charge density & $(n,\nu_1,\nu_2,\nu_3)$ & $(n,-\nu_1,-\nu_2,-\nu_3)$  \\
    \hline
    Momentum density & $(\kappa_1,\kappa_2,\tau_3,\tau_4)$ & $(-\theta_2,-\theta_1,-\theta_3, 0)$  \\
    Charge flux & $(\tau_1,\tau_2,\chi_1,\chi_2)$ & $(-\gamma_2 T,-\gamma_1 T,-\gamma_3 T ,0)$  \\
    Shear viscosity & $\eta$ & $-\eta$  \\
    \hline
    Frame invariants & $\left(\tilde \lambda, \tilde \zeta, \tilde \nu,\tilde \tau_1,\tilde \tau_2,\tilde \chi_1,\tilde \chi_2  \right)$ & $(-f_1,-f_2,-f_3,-\ell_2 T,-\ell_1 T,-\ell_3 T,0)$ \\
    Frame invariants~(on-shell) & $\left({\bm \zeta}, {\bm \chi}_1, {\bm \chi}_2, {\bm \tau} \right)$ & $(\zeta, -T^2 \sigma,{\rm none},T^2 \chi_{\rm T})$ \\
    \hline
  \end{tabular}
\end{table}

\section{Useful equations}
\label{app:usfleqs}

\subsection{Thermodynamical relations}

In the main text, we often use the following relations:
\begin{itemize}
	\item The first law of thermodynamics:
	\begin{align}
		 \rho_\beta -\mu n_\beta =\frac{s_\beta}{\beta},~\rho_\mu - \mu n_\mu = \frac{s_\mu}{\beta}.
\end{align}
	\item Thermodynamical relations:
	\begin{align}
	P_\beta & = -\frac{s}{\beta^2},~P_\mu = n,\label{thr}
\end{align}
\item
	The integrability condition:
\begin{align}
	-\frac{s_\mu}{\beta^2}   =n_\beta.\label{ibc}
\end{align}
\end{itemize}

\subsection{KMS condition}
\label{KMSsummary}

For readers' convenience, we present the most general thermodynamical first-order constitutive relations after imposing the FDR:
\begin{align}
	\rho_{(1)} &= \beta^{-1}\epsilon u^\rho \nabla_\rho \beta  - \lambda \nabla_\rho u^\rho - \nu_1 u^\rho \nabla_\rho (\beta\mu)  ,
	\label{thmconsbegin}
	\\
	P_{(1)} &= \beta^{-1} \lambda u^\rho \nabla_\rho \beta  - \zeta \nabla_\rho u^\rho  -  \nu_2 u^\rho \nabla_\rho (\beta \mu) ,
	\\
	n_{(1)} & =  \nu_1 u^\rho \nabla_\rho  \beta  - \nu_2 \beta \nabla_\rho u^\rho -  \nu_3 \beta u^\rho \nabla_\rho(\beta   \mu),
\\
	q_{(1)\mu} &= \kappa (\beta^{-1} \nabla_\mu  \beta -   u^\rho\nabla_\rho  u_\mu ) - \tau (\nabla_\mu (\beta \mu) + \beta  u^\rho  F_{\rho\mu}),
	\\
	j_{(1)\mu} &= \tau( \nabla_\mu  \beta -   \beta u^\rho\nabla_\rho  u_\mu)  -  \chi ( \beta  \nabla_\rho  (\beta \mu) + \beta^2  u^\rho F_{\rho\mu}),
\\
	\begin{split}
		\beta \hbar W_{(0)}^{\mu\nu\rho\sigma} 
	&=\epsilon u^\mu u^\nu u^\rho u^\sigma 
	+ \lambda ( \gamma^{\mu \nu}u^\rho u^\sigma + \gamma^{\rho\sigma}u^\mu u^\nu )
	\\
	&+ \kappa ( \gamma^{\mu \rho}u^\nu u^\sigma + \gamma^{\nu \rho}u^\mu u^\sigma +\gamma^{\mu \sigma}u^\nu u^\rho+\gamma^{\nu \sigma}u^\mu u^\rho)
	\\ 
	&+ \zeta \gamma^{\mu\nu}\gamma^{\rho\sigma}
	+ \eta\left( \gamma^{\mu\rho}\gamma^{\nu\sigma} 
	+\gamma^{\nu\rho}\gamma^{\mu\sigma} 
	-\frac{2}{3}\gamma^{\mu\nu}\gamma^{\rho\sigma}\right ),
\end{split}
\label{def:W0}
\\
\begin{split}
		\beta \hbar X_{(0)}^{\mu\nu\rho} 
	&=\nu_1 u^\mu u^\nu u^\rho  
	+ \nu_2  \gamma^{\mu \nu}u^\rho  
	+ \tau ( \gamma^{\mu \rho}u^\nu  + \gamma^{\nu \rho}u^\mu  ),
\end{split}
\label{def:Wd}
\\
\begin{split}
		\beta \hbar Y_{(0)}^{\mu \rho } 
	&=\nu_3 u^\mu  u^\rho  
	+ \chi  \gamma^{\mu \rho}.
\end{split}
\label{thmconsend}
\end{align}

\subsection{Adiabatic sound velocity}\label{appadsv}

The adiabatic sound velocity in terms of thermodynamical variables is found as follows.
By differentiating $P=P(\rho,s/n)$, we get 
\begin{align}
	dP &= \left( \frac{\partial P }{\partial \rho} \right)_{s/n} d\rho + \left( \frac{\partial P}{\partial (s/n)} \right)_{\rho} d (s/n)
		\notag 
	\\
	&= \left( \frac{\partial P }{\partial \rho} \right)_{s/n} \left( \rho_\mu d\mu + \rho_\beta d\beta \right)+ \frac{s}{n}\left( \frac{\partial P}{\partial (s/n) } \right)_{\rho}\left(\frac{s_\mu d\mu + s_\beta d\beta}{s}-\frac{n_\mu d\mu + n_\beta d\beta }{n}\right).
\end{align}
Combining this with $dP = P_\beta d\beta + P_\mu d\mu$, one finds
\begin{align}
		\left( \frac{\partial P}{\partial \rho} \right)_{s/n}&=\frac{n (P_\mu  \rho_\beta -P_\beta  \rho_\mu )+( n_\mu P_\beta  -n_\beta P_\mu)  (\rho +P) }{(\rho +P) (n_\mu \rho_\beta -n_\beta \rho_\mu )} = c_{s/n}^2.\label{exp:adiabaticsv}
\end{align}
Similarly, the speeds of iso-entropic sound waves and iso-number sound waves are
\begin{align}
	\left(\frac{\partial P}{\partial \rho}\right)_s &= \frac{ \mu  ( n_\mu P_\beta-  n_\beta P_\mu )  +P_\mu  \rho_\beta -P_\beta  \rho_\mu  }{\mu  ( n_\mu \rho_\beta -  n_\beta \rho_\mu) } = c^2_s , \label{prs}
	\\
	\left(\frac{\partial P}{\partial \rho}\right)_n &= \frac{n_\mu P_\beta -n_\beta P_\mu }{n_\mu \rho_\beta -n_\beta \rho_\mu } = c_n^2. \label{prn}
\end{align}
These sound velocities are related as
\begin{align}
	c_{s/n}^2 = \frac{n c_s^2 + s c_n^2}{s+\beta \mu n}= \frac{n c_s^2 + s c_n^2}{\rho+ P }.
\end{align}
Also, the differentiation of $P$ with respect to $s$ and $n$ are written as
\begin{align}
	\left(\frac{\partial P}{\partial n}\right)_\rho &= \frac{P_\mu  \rho_\beta -P_\beta  \rho_\mu }{n_\mu \rho_\beta -n_\beta \rho_\mu }  ,\label{pnr}
	\\
	\left(\frac{\partial P}{\partial s}\right)_\rho &= -\frac{P_\mu  \rho_\beta -P_\beta  \rho_\mu }{\beta  \mu  ( n_\mu \rho_\beta -  n_\beta \rho_\mu )} .\label{psr}
\end{align}

\section{1D Brownian motion}
\label{appBM}

There are many similarities between the dynamics of the NG modes in fluids and a Brownian particle $X$ bounded in a harmonic potential.
For the reader's convenience, we provide a quick review of a harmonic oscillator in a thermal environment.
Let $m$ be the mass of a Brownian particle, $\omega_0$ be the frequency of the oscillator, and $\gamma$ be the viscosity coefficient.
The equation of motion for a 1-dimensional harmonic oscillator $X$ in a thermal environment of the inverse temperature $\beta$ is
\begin{align}
	m\ddot X + m\gamma \dot X + m \omega^2_0 X = \xi,~\langle \xi(t)\xi(t')\rangle = \frac{2 m \gamma}{\beta} \delta(t-t').\label{BMeom}
\end{align}
Using the Green function $G$ such that
\begin{align}
	\ddot G + \gamma \dot G + \omega^2_0 G = \delta(t-t'),
\end{align}
we find 
\begin{align}
	X = \frac{1}{m}\int_0^t d t' G(t- t') \xi(t').\label{intXG}
\end{align}
The Keldysh propagator is found as
\begin{align}
	\langle X(t)X(t_1)\rangle =  \frac{2\gamma}{m\beta}\int_0^t dt'\int_0^{t_1} dt'_1 G(t-t')G(t_1-t'_1) \delta(t' - t'_1). 
\end{align}
For simplicity, consider $t\geq t_1$.
Then
\begin{align}
	\langle X(t)X(t_1)\rangle &=  \frac{2\gamma}{m\beta}\int_0^t dt' G(t-t')G(t_1-t')\Theta(t_1-t')
	\notag 
	\\
	&=  \frac{2\gamma}{m\beta}\Theta(t-t_1)\int_0^{t_1} dt' G(t-t')G(t_1-t').\label{eqcordef} 
\end{align}
$t\leq t_1$ is similarly evaluated, and we find the unequal time correlation functions for $X(t)$.

\medskip
$G$ is found as follows.
First, we consider the Fourier integral of $G$:
\begin{align}
	G(t-t') = \int \frac{d\omega}{2\pi}e^{-i\omega(t-t')}G(\omega) 
\end{align}
Then, the EoM in Fourier space is
\begin{align}
	-\omega^2  G(\omega) - i\omega \gamma  G(\omega) + \omega^2_0 G(\omega) = 1,
\end{align}
which yields
\begin{align}
	G(\omega) = \frac{1}{\omega_0^2-\omega^2 - i\omega \gamma}.
\end{align}
Fourier transforming $G(\omega)$, we find
\begin{align}
	G(t) = \int_{-\infty}^{\infty} \frac{d\omega}{2\pi}\frac{e^{-i \omega t}}{\omega_0^2-\omega^2 - i\omega \gamma} = \Theta(t)\frac{e^{-\frac{\gamma t}{2}}}{\omega_0}\frac{\sin\left( \omega_0t\sqrt{1 - \frac{\gamma^2}{4 \omega_0^2}}\right)}{\sqrt{1 - \frac{\gamma^2}{4 \omega_0^2}}}\label{gflgb}.
\end{align}

\medskip
Finally, let us evaluate Eqs.~\eqref{eqcordef} by using \eqref{gflgb}.
We are particularly interested in the case where $(t+t_1)/2\to \infty$ while $t-t_1$ is finite: sufficient time has passed since the initial time of the thermal state, but consider the correlation for a finite time interval.
In this limit, the analytic expression simplifies, and in the small $\gamma$ limit, we find the following asymptotic form:
\begin{align}
	\langle X(t)X(t_1)\rangle &\simeq  \frac{e^{-\frac{1}{2} \gamma |t-t_1|}}{m\beta \omega_0^2} \left(\cos \left( \omega_0 |t-t_1| \sqrt{1-\frac{\gamma^2}{4\omega_0^2}}\right) + \frac{\gamma}{2\omega_0 } \frac{\sin \left(\omega_0 |t-t_1| \sqrt{1-\frac{\gamma^2}{4\omega_0^2}} \right)}{\sqrt{1-\frac{\gamma^2}{4\omega_0^2}}}\right).\label{gktt'}
\end{align}
Thus, $t=t_1$ is fully correlated, and then the correlation is exponentially suppressed.

\medskip
Let us find the same expression from the path integral.
The MSR action for Eq.~\eqref{BMeom} is written as
\begin{align}
	iS = \int dt \left[ -i (m\ddot X + m\gamma \dot X + m \omega^2_0 X){\bar X} - \frac{1}{2} \frac{2m\gamma}{\beta} {\bar X}^2  \right].\label{msrbp}
\end{align}
For notational simplicity, we introduce the canonically normalized variables
\begin{align}
	y = \sqrt{m} X,~ {\bar y} = \sqrt{m} {\bar X},
\end{align}
and then
\begin{align}
	iS &= \int dt \left[ -i (\ddot y + \gamma \dot y + \omega^2_0 y){\bar y} -  \frac{\gamma}{\beta} {\bar y}^2  + i\bar j y +i j {\bar y}\right],
\end{align}
where $j$ and $\bar j$ are the sources.
Consider the action in Fourier space. 
Using
\begin{align}
	 \int dt A(t)B(t) = \int dt \int \frac{d\omega'}{2\pi} e^{i\omega' t}A_{\omega'}\int \frac{d\omega}{2\pi} e^{i\omega t}B_{\omega}=  \int \frac{d\omega}{2\pi } A^*_\omega B_{\omega},
\end{align}
and
\begin{align}
	Y = \binom{y}{{\bar y}},~J = \binom{\bar j}{j},~B = 	\begin{pmatrix}
		0 & \omega^2 - \omega^2_0 - i\omega \gamma 
		\\
		\omega^2 - \omega^2_0 + i\omega \gamma  &    \frac{2i\gamma}{\beta}
	\end{pmatrix}
\end{align}
the action is recast into
\begin{align}
	iS &= \frac{i}{2}\int \frac{d\omega}{2\pi} 
	Y^\dagger B Y
	+ J^\dagger Y + Y^\dagger J,
\end{align}
where $J^\dagger Y=Y^\dagger J$.
In the standard prescription for zero temperature path integral, we consider a functional shift 
\begin{align}
	Y \to Y - B^{-1} J, ~iB^{-1} = 
	\begin{pmatrix}
		\frac{2 \gamma}{\beta \left(\gamma^2 \omega ^2+\left(\omega ^2-\omega_0^2\right)^2\right)} & \frac{1}{\gamma \omega -i \left(\omega ^2-\omega_0^2\right)}
		\\
		-\frac{1}{\gamma \omega +i \left(\omega ^2-\omega_0^2\right)} & 0
	\end{pmatrix}
	\label{funcshift}
\end{align}
which eliminates the cross-term of $J$ and $Y$.
In the present case, $B$ is not Hermitian due to the noise term.
In fact, we find
\begin{align} 
	&(Y^\dagger - J^\dagger (B^{-1})^\dagger) B (Y - B^{-1}J)
	+ J^\dagger (Y - B^{-1}J) + (Y^\dagger - J^\dagger (B^{-1})^\dagger) J
	\notag 
	\\
	=&
	Y^\dagger BY - J^\dagger (B^{-1})^\dagger  BY - Y^\dagger B B^{-1}J + J^\dagger (B^{-1})^\dagger  B B^{-1}J  
	\notag 
	\\
	&
	+ J^\dagger Y - J^\dagger B^{-1}J + Y^\dagger J- J^\dagger (B^{-1})^\dagger J
	\notag 
	\\
	=&
	Y^\dagger BY - J^\dagger (B^{-1})^\dagger  BY  	+ J^\dagger Y - J^\dagger B^{-1}J. 
\end{align}
For a Hermitian $B$, we get $(B^{-1})^\dagger  B  = 1$, and then $J^\dagger Y$ is subtracted.
In our case, $B$ is not Hermitian: $(B^{-1})^\dagger  B\neq 1$.
We are not able to integrate $y$ independently from $j$ due to this nonvanishing cross-term.
In fact, Eq.~\eqref{funcshift} is not a unique way to find the bilinear form of $J$.
Instead, one may consider $Y^\dagger \to Y^\dagger - Y^\dagger B^{-1}$.
Then we find $Y^\dagger J - Y^\dagger B (B^{-1})^\dagger J$ remains.
These terms are $\gamma/\beta$ multiplied by the retarded and advanced propagators, respectively.
Then, we may split the path integral for $t<0$ and $t>0$ and eliminate these cross terms.
Finally, we find the Green functions found in Eqs.~\eqref{gflgb} and~\eqref{gktt'} after rescaling by $m$:
\begin{align}
	\int \frac{d\omega}{2\pi}e^{-i\omega t}	
	iB^{-1} =
	\begin{pmatrix}
		G_K 
		&
		G_R
		\\
		G_A
		&
		0		
	\end{pmatrix},
\end{align}
where $G_R=iG$.

\bibliography{biblio.bib}{}
\bibliographystyle{unsrturl}

\end{document}